\documentclass[preprint,12pt]{elsarticle}

\usepackage[outdir=./plots/epstopdf/]{epstopdf}
\usepackage{bm}
\usepackage{amsmath}
\usepackage{graphicx}
\usepackage{subcaption}
\usepackage{multirow}
\usepackage{makecell}
\usepackage{pdflscape}
\usepackage{tabularx}
\usepackage{overpic}
\usepackage{placeins}
\usepackage{caption}
\usepackage[ruled,vlined]{algorithm2e}
\usepackage[justification=centering]{caption}
\usepackage{algpseudocode}
\usepackage{xcolor}

\usepackage{amssymb}
\usepackage{amsthm}

\DeclareMathOperator*{\argmin}{arg\,min}
\DeclareMathOperator*{\argmax}{arg\,max}

\newtheorem{theor}{Theorem}[section]
\newtheorem{remark}{Remark}[section]
\newtheorem{assum}{Assumption}

\biboptions{sort&compress}

\journal{TBD}

\begin{document}

\begin{frontmatter}



\title{Bayesian improved cross entropy method for network reliability assessment}

\author{Jianpeng Chan}
\author{Iason Papaioannou}
\author{Daniel Straub}
\address{Engineering Risk Analysis Group, Technische Universit{\"a}t M{\"u}nchen, Arcisstr. 21, 80290 M{\"u}nchen, Germany}
 
\begin{abstract}
We propose a modification of the improved cross entropy (iCE) method to enhance its performance for network reliability assessment. The iCE method performs a transition from the nominal density to the optimal importance sampling (IS) density via a parametric distribution model whose cross entropy with the optimal IS is minimized. The efficiency and accuracy of the iCE method are largely influenced by the choice of the parametric model. In the context of reliability of systems with independent multi-state components, the obvious choice of the parametric family is the categorical distribution. When updating this distribution model with standard iCE, the probability assigned to a certain category often converges to 0 due to lack of occurrence of samples from this category during the adaptive sampling process, resulting in a poor IS estima tor with a strong negative bias. To circumvent this issue, we propose an algorithm termed Bayesian improved cross entropy method (BiCE). Thereby, the posterior predictive distribution is employed to update the parametric model instead of the weighted maximum likelihood estimation approach employed in the original iCE method. A set of numerical examples illustrate the efficiency and accuracy of the proposed method.      

\end{abstract}


\end{frontmatter}

\graphicspath{{figures/}{tikz/}{plots/}{epstopdf/}}


\section{Introduction}
\label{sec: Intro}
Infrastructure networks, such as power transmission networks and water supply systems, operate as backbones of urban communities. Hence, it is essential to properly quantify the risk of network failure, which involves quantification of the failure probability of the network system. A network is considered as failed when it cannot deliver a specified level of performance. Mathematically, the failure is described through a function $g(\cdot)$, known as performance function, structure function, or limit state function (LSF). 
Let $\bm{X}$ be a $n$-dimensional vector of random variables with joint density function $p_{\bm{X}}(\bm{x})$.
The failure event $F$ collects all system states $\bm{x}$ whose LSF $g(\bm{x})$ is less or equal than 0, i.e., $F\triangleq \{\bm{x}: g(\bm{x}) \leq 0 \}$. The probability of failure is defined as
\begin{equation}
\label{Eq: Failure probability}
p_f \triangleq \mathbb{P}(g(\bm{X})\leq0) = \mathbb{E}_p[\mathbb{I}\{g(\bm{X})\leq 0\}],
\end{equation}
where $\mathbb{I}\{\cdot\}$ represents the indicator function, and $\mathbb{E}_p$ denotes expectation with respect to the density $p_{\bm{X}}(\bm{x})$.\\ 
The network performance is often measured through connectivity or flow. In connectivity-based problems, one evaluates the probability that a given set of nodes are disconnected \cite{Li&Liu2021, Ball&others1995}, and typically, both the network performance and the component state are modeled as binary random variables. 
In flow-based problems, one is interested in the flow that a network can deliver, e.g., the maximum number of passengers that can be transported from one city to another through the railway network. Multi-state or continuous random variables are often involved in this class of problems. For water supply systems and power grids, the flow is driven by the physical law and operation strategies, and the initial failure of network components leads to a reconfiguration of the power flow that may trigger additional cascading failure.\\ 
A set of methodologies have been proposed for evaluating the reliability in the above two classes of problems, among which sampling-based methods such as Monte Carlo simulation (MCS) and its different variants feature prominently \cite{Kumamoto&others1980, Fishman1986a, Fishman1989, Alexopoulos&Fishman1991, Elperin&others1992, Cancela&Khadiri2003, Rubino&Tuffin2009, Zio2013, Behrensdorf&others2021}. 
For rare event simulation, i.e., when the failure probability $p_f$ is small, crude MCS is inefficient or even infeasible when the LSF is expensive to compute. In such cases, advanced sampling techniques such as subset simulation \cite{Zio&Pedroni2008, Botev&others2013, Zuev&others2015, Botev&others2018, Jensen&Jerez2018, Chan&others2022a} and importance sampling (IS) \cite{Bulteau&Khadiri2002, Hui&others2003, Wang&others2016} should be employed to decrease the required number of LSF evaluations for obtaining an accurate estimate of $p_f$. 
Alternatively, Dehghani et. al. \cite{Dehghani&others2021} employ an actively trained surrogate model to approximate the computational demanded LSF, resulting in an accurate and efficient estimator.\\ 
In this paper, we focus on the IS technique for rare event estimation in static (or time independent) network reliability problems.   
The performance of the IS method strongly depends on the choice of the IS density. The cross entropy (CE) method \cite{Rubinstein1997} determines the IS density as the member of a parametric family that has the minimal Kullback-Leibler (KL) divergence/distance from the optimal IS density $p^*_{\bm{X}}(\bm{x})$. For rare event estimation, the KL divergence is minimized iteratively between the parametric family and a sequence of intermediate target densities that gradually approach $p^*_{\bm{X}}(\bm{x})$. The resulting density in the last iteration is then employed as the IS density for estimating $p_f$. Papaioannou et. al \cite{Papaioannou&others2019} further enhanced the performance of the CE method through modifying the intermediate target densities. The basic idea of this approach, termed improved cross entropy (iCE) method, is to introduce a smooth transition from the input density to the optimal IS density to make better use of the intermediate samples.\\ 
In the context of network reliability assessment with discrete multi-state components, the obvious choice of the parametric family is the categorical distribution. However, updating the categorical model with the CE or iCE method can perform poorly, especially when the sample size is small. This is because the probability assigned to a certain category often converges to 0 when no samples fall into this category during the adaptive sampling process. This is known in the literature as the {\it zero count problem} \cite{Murphy2012}. Neglecting a certain category in the IS distribution can lead to a bias in the IS estimate of $p_f$. 
To avoid such issue, one may think of transferring the discrete random variable space to a continuous one through, for example, the Rosenblatt transformation \cite{Chan&others2022a} and employ continuous parametric families in the iCE method. However the network reliability problem becomes more challenging after this non-linear transformation and the iCE method often fails to converge.
Hui et. al. \cite{Hui&others2005} combine the cross entropy method with the graph creation process \cite{Elperin&others1991} and efficiently estimate the connectivity reliability of networks using an independent exponential parametric model. Note that this method is computational demanded and applies only to coherent binary systems. In this paper, we employ the independent categorical distribution as IS distribution and propose an approach for learning its parameters during the iCE sampling process that avoids the zero count problem. The proposed algorithm, termed Bayesian improve cross entropy method (BiCE), employs the posterior predictive distribution to update the parametric family instead of the weighted maximum likelihood estimator used in the standard CE method. 
Compared with other non-sampling based methods (e.g., \cite{Li&He2002, Hardy&others2007, Paredes&others2019, Miao&others2020, Byun&Song2021}), the proposed BiCE method facilitates using advanced network analysis algorithms that account for complex network dynamics. However, the BiCE may requires a large number of samples to achieve an acceptable results.\\  
The rest of the paper is organized as follows. A brief introduction to the IS approach is given in Section 2. In Section 3, we review the CE and iCE methods and provide some new insights into these two methods. In Section 4, we first illustrate the problem that occurs when updating the categorical distribution using CE or iCE, and then propose the BiCE method to circumvent this problem. A set of numerical samples is given in Section 5 to illustrate the efficiency and accuracy of the proposed approach.
\section{Importance sampling}
\label{sec: intro to IS}
Estimation of $p_f$ in Eq.(\ref{Eq: Failure probability}) using crude MCS is straightforward; one generates $N$ samples from the joint density function $p_{\bm{X}}(\bm{x})$ and then take the sample mean of the indicator function as the unbiased estimator of $p_f$. The coefficient of variation (c.o.v.) of the MCS estimate equals $\sqrt{ \frac{1-p_f}{N\cdot p_f} }$; therefore, for small $p_f$ the required number of samples for achieving an accurate result is large. For rare event estimation, acceleration techniques to speed up the occurrence of the failure events are necessary. IS is an efficient and widely utilized method for efficient simulation of rare events.\\
The basic idea of IS is to sample from a proposal distribution, also known as IS distribution, under which the rare event is more likely to occur and to correct the resulting bias in the estimate by multiplying each sample in the IS estimator with an appropriate likelihood ratio $L$ \cite{Rubino&Tuffin2009}. Specifically, let $p_{IS}(\bm{x})$ denote the IS density and $\{\bm{x}_{k}\}_{k=1}^{N}$ be the $N$ samples generated from $p_{IS}(\bm{x})$. The IS estimator of the failure probability in Eq.(\ref{Eq: Failure probability}) reads
\begin{equation}
\label{Eq: IS estimator}
\widehat{p}_f = \frac{1}{N}\sum_{k=1}^N \mathbb{I}\{ g(\bm{x}_k)\leq 0 \} \frac{p_{\bm{X}}(\bm{x}_k)}{p_{IS}(\bm{x}_k)},
\end{equation}
where the likelihood ratio (or IS weight) $L(\bm{x}) \triangleq \frac{p_{\bm{X}}(\bm{x})}{p_{IS}(\bm{x})}$ can be interpreted as an adjustment factor that compensates for the fact that samples are generated from $p_{IS}(\bm{x})$ instead of $p_{\bm{X}}(\bm{x})$ \cite{Owen2013}. The IS estimator in Eq.(\ref{Eq: IS estimator}) is unbiased if the failure domain $F$ is included in the sample space of $p_{IS}(\bm{x})$ \cite{Owen2013}. The variance of the estimator mainly depends on the choice of the IS distribution. A proper choice of the IS distribution can lead to significantly smaller variance than that of crude MCS. Indeed,  the theoretical optimal IS distribution $p^*_{\bm{X}}(\bm{x})$ that results in zero variance of the estimator is equal to the input distribution conditional on occurrence of the failure event. That is   
\begin{equation}
\label{Eq: optimal IS distribution}
p^*_{\bm{X}}(\bm{x}) = \frac{p_{\bm{X}}(\bm{x})\mathbb{I}\{g(\bm{x})\leq0 \}}{p_f} = p_{\bm{X}}(\bm{x}|F).
\end{equation}
Unfortunately, $p^*_{\bm{X}}(\bm{x})$ cannot be directly used, since its analytical expression relies on a prior knowledge of the sought failure probability $p_f$. Nevertheless, the optimal IS distribution $p^*_{\bm{X}}(\bm{x})$ still provides guidance for selecting an appropriate IS distribution. A common approach is to perform an initial first/second order reliability method analysis \cite{Madsen&others2006} or employ a Markov chain simulation algorithm \cite{Au&Beck1999} to form a distribution that resembles $p^*_{\bm{X}}(\bm{x})$. Alternatively, one can approximate $p^*_{\bm{X}}(\bm{x})$ in a adaptive manner through application of the CE or iCE methods, which are discussed in detailed in Section 3.
\section{Cross entropy and improved cross entropy method}
\label{sec: intro to CE and iCE}
\subsection{Cross entropy method}
The CE method determines the IS distribution in the estimator in Eq.(\ref{Eq: IS estimator}) through minimizing the KL divergence between the theoretical optimal IS distribution $p^*_{\bm{X}}(\bm{x})$ and a predefined parametric family of distributions. The KL divergence, which is also known as relative entropy, is a measure of how one distribution differs from another. Specifically, let $h(\bm{x}; \bm{v})$ denote a family of parametric distributions, where $\bm{v} \in \mathcal{V}$ is a parameter vector. The KL divergence between $p^*_{\bm{X}}(\bm{x})$ and $h(\bm{x}; \bm{v})$ is defined as \cite{Rubinstein&Kroese2016}
\begin{align}
D( p^*_{\bm{X}}, h )&= \mathbb{E}_{p^*_{\bm{X}}} \left[ \ln \left( \frac{p^*_{\bm{X}}(\bm{X})}{ h(\bm{X}; \bm{v}) } \right) \right] \notag \\
\label{Eq: KL}&=\mathbb{E}_{p^*_{\bm{X}}}[\ln(p^*_{\bm{X}}(\bm{X}))]-\mathbb{E}_{p^*_{\bm{X}}}[\ln(h(\bm{X}; \bm{v}))].
\end{align}
In order to obtain a precise IS estimator, the KL divergence $D( p^*_{\bm{X}}, h )$ needs to be small. In fact, one can prove that \cite{Au&Beck2003} the c.o.v. of the IS estimator, $\delta(\widehat{P}_f)$ is lower-bounded by 
\begin{equation}
\label{Eq: c.o.v. and KL}
\delta(\widehat{P_f}) \geq \sqrt{\frac{\text{exp}( D(p^*_{\bm{X}}, h))-1}{N}},
\end{equation}
According to Eq.(\ref{Eq: c.o.v. and KL}), if we require that $\delta(\widehat{P_f}) \leq 0.1$, the KL divergence $ D( p^*_{\bm{X}}, h )$ should be less or equal than $ \ln(1+0.01N)$. Conversely, a large KL divergence leads to a high c.o.v. and hence an imprecise result.\\ 
The CE method determines the optimal parameter vector $\bm{v}^*$ through minimizing the KL divergence of Eq.(\ref{Eq: KL}), i.e., through solving
\begin{equation}
\label{Eq: CE optimization prob.1}
\bm{v}^* = \argmin\limits_{\bm{v}\in\mathcal{V}}D(p^*_{\bm{X}}, h).
\end{equation}
Since the first term on the right hand side of Eq.(\ref{Eq: KL}) does not depend on $\bm{v}$, Eq.(\ref{Eq: CE optimization prob.1}) is equivalent to
\begin{equation}
\label{Eq: CE optimization prob.2}
\bm{v}^* = \argmin\limits_{\bm{v}\in\mathcal{V}}-\mathbb{E}_{p^*_{\bm{X}}}[\ln(h(\bm{X}; \bm{v}))].
\end{equation}
Typically, the optimization problem in Eq.(\ref{Eq: CE optimization prob.2}) is convex and can be solved by the Lagrange multiplier method \cite{Boyd&Vandenberghe2004}. However, the objective function depends on the optimal IS distribution $p^*_{\bm{X}}(\bm{x})$, which is not known in closed form, and therefore Eq.(\ref{Eq: CE optimization prob.2}) cannot be solved analytically. Instead, we estimate $\bm{v}^*$ through solving an alternative objective function, which is introduced in the following.\\ 
Substituting $p^*_{\bm{X}}$ in Eq.(\ref{Eq: CE optimization prob.2}) with the expression of Eq.(\ref{Eq: optimal IS distribution}), one gets
\begin{equation}
\label{Eq: CE optimization prob.3}
\bm{v}^* = \argmax\limits_{\bm{v}\in\mathcal{V}} 
\mathbb{E}_p[ \mathbb{I} \{ g(\bm{X})\leq0 \} \ln(h(\bm{X}; \bm{v})) ]
\end{equation}
The expectation in Eq.(\ref{Eq: CE optimization prob.3}) can be approximated through IS, which gives the importance sampling counterpart of the CE  optimization problem. That is  
\begin{equation}
\label{Eq: samp. counterpart of the CE optimization prob.}
\widehat{\bm{v}} = \argmax\limits_{\bm{v}\in\mathcal{V}} \frac{1}{N} \sum \limits_{k=1}^{N} \frac{p_{\bm{X}}(\bm{x}_k)\mathbb{I}\{g(\bm{x}_k)\leq 0\}}{p_{ref}(\bm{x}_k)}\ln(h(\bm{x}_k; \bm{v})), \quad\quad \bm{x}_k \sim p_{ref}(\cdot).
\end{equation}
Here, $p_{ref}(\bm{x})$ is the IS distribution used to estimate the expectation in Eq.(\ref{Eq: CE optimization prob.3}) and is termed the reference distribution in the CE method \cite{Rubinstein&Kroese2016}. Similarly to the original CE optimization problem, the optimization problem in Eq.(\ref{Eq: samp. counterpart of the CE optimization prob.}) can also be solved by the Lagrange multiplier method.\\
One should distinguish $h(\bm{x}; \bm{v}^*)$ from $h(\bm{x};\widehat{\bm{v}})$ in the CE method \cite{Chan&Kroese2012}. $h(\bm{x}; \bm{v}^*)$ represents the distribution that has the smallest KL divergence $D( p^*_{\bm{X}}, h )$ among a set of distributions and hence is termed the sub-optimal IS distribution. $h(\bm{x};\widehat{\bm{v}})$ is the distribution we use as the IS distribution, i.e., the distribution resulting from solution of the optimization problem of Eq. (\ref{Eq: samp. counterpart of the CE optimization prob.}). We term it the chosen IS distribution for the rest of the paper. Note that, as long as the parametric family is fixed, the 'distance' between the optimal IS distribution and the sub-optimal IS distribution is also fixed. The objective of the CE method is finding a good estimator $\widehat{ \bm{v} }$ that is close to the optimal but inaccessible CE parameter $\bm{v}^*$.
\begin{remark}
\label{Remark 1}
In general, if $h(\bm{x}; \bm{v})$ is a properly parameterized exponential family, $\widehat{\bm{v}}$ can be interpreted as the self-normalized IS estimator of $\bm{v}^*$. The accuracy of the self-normalized IS estimator is measured by the effective sample size (ESS). For more details we refer to \ref{Appendix: Self-normalized importance sampling and cross entropy method}.
\end{remark}
\begin{remark}
\label{Remark 2}
$\widehat{\bm{v}}$ can also be interpreted as a weighted maximum likelihood estimation (MLE) of $\bm{v}$ \cite{Geyer&others2019} and therefore may suffer from the same drawbacks as MLE (e.g., overfitting). To circumvent the overfitting issue of $\widehat{\bm{v}}$, we propose a novel Bayesian estimator $\widetilde{\bm{v}}$ for the CE method in Section \ref{sec: Categorical parametric model}. The proposed estimator converges to $\bm{v}^*$ as the sample size goes to infinity.
\end{remark}

\subsection{Cross entropy method for rare events and improved cross entropy method}
The efficiency and accuracy of the CE method depend on the choice of the reference distribution $p_{ref}(\bm{x})$ in Eq.(\ref{Eq: samp. counterpart of the CE optimization prob.}). A potential choice for $p_{ref}(\bm{x})$ is the input distribution $p_{\bm{X}}(\bm{x})$. However, for the case where $F=\{\bm{x}: g(\bm{x})\leq0\}$ is a rare event, sampling directly from $p_{\bm{X}}(\bm{x})$ will lead to a large number of zero indicators in Eq.(\ref{Eq: samp. counterpart of the CE optimization prob.}), and, hence, an inaccurate result.\\ 
In such case, the reference distribution can be chosen adaptively. Let $p^{(t)}(\bm{x}), t=1,...,T$ denote a sequence of intermediate target distributions that gradually approach the optimal IS distribution $p^*_{\bm{X}}(\bm{x})$. The CE optimization problem is then solved iteratively by finding a good approximation to each intermediate target distribution, resulting in a sequence of CE parameter vectors $\{\widehat{\bm{v}}^{(t)}, t=1,...,T \}$ and distributions $\{h(\bm{x}; \widehat{\bm{v}}^{(t)}), t=1,...,T \}$. The distribution one obtains in the $t$-th iteration, $h(\bm{x}; \widehat{\bm{v}}^{(t)})$, is used as the reference distribution $p_{ref}(\bm{x})$ for the CE procedure in iteration $t+1$. For the first iteration, the input distribution $p_{\bm{X}}(\bm{x})$ is used as the reference distribution. In this way, one takes $h(\bm{x}; \widehat{\bm{v}}^{(T-1)})$ as the reference distribution $p_{ref}(\bm{x})$ for Eq.(\ref{Eq: samp. counterpart of the CE optimization prob.}), and $h(\bm{x}; \widehat{\bm{v}}^{(T)})$ as the final IS distribution. The goal is to make $\widehat{\bm{v}}^{(T)}$ a good estimator of $\bm{v}^*$. Typically, the intermediate target distributions $p^{(t)}(\bm{x})$ are not predefined but are chosen adaptively during the iterations. Depending on the way of adaptively selecting $p^{(t)}(\bm{x})$, one distinguishes the (multilevel) CE method and its improved version, the improved cross entropy (iCE) method.\\
For the CE method, the intermediate target distributions are defined as:
\begin{equation}
\label{Eq: intermediate target distribution of CE method}
p^{(t)}(\bm{x}) \triangleq \frac{1}{Z^{(t)}} p_{\bm{X}}(\bm{x})\mathbb{I}\{g(\bm{x})\leq \gamma^{(t)} \}, t=1,...,T
\end{equation}   
where $\{\gamma^{(t)}, t=1,....T\}$ is a parameter vector that satisfies $\gamma^{(t)} \geq 0$, and $Z^{(t)}$ is a normalizing constant. The CE optimization problem for Eq.(\ref{Eq: intermediate target distribution of CE method}) reads
\begin{equation}
\label{Eq: CE optimization for intermediate target dist. in CE}
\bm{v}^{(t,*)} = \argmax\limits_{\bm{v}\in\mathcal{V}} 
\mathbb{E}_p[ \mathbb{I}\{ g(\bm{X}) \leq \gamma^{(t)} \} \ln(h(\bm{X}; \bm{v})) ].
\end{equation}
The sample counterpart of the CE optimization problem for Eq.(\ref{Eq: CE optimization for intermediate target dist. in CE}) reads as follows:
\begin{equation}
\label{Eq: samp. counterpart for intermediate target dist. in CE}
\widehat{\bm{v}}^{(t)} = \argmax\limits_{\bm{v}\in\mathcal{V}} \frac{1}{N} \sum \limits_{k=1}^{N} \frac{p_{\bm{X}}(\bm{x}_k)\mathbb{I}\{g(\bm{x}_k)\leq \gamma^{(t)} \}}{p_{ref}(\bm{x}_k)}\ln(h(\bm{x}_k; \bm{v})), \quad \bm{x}_k \sim p_{ref}(\cdot).
\end{equation}
In the $t$-th iteration, the CE method proceeds through the following three steps:
(1) Generate a set of samples $\mathcal{P}^{(t)} \triangleq \{\bm{x}_k, k=1,...,N\}$ from the reference distribution $p_{ref}(\bm{x})=p_{\bm{X}}(\bm{x})$ in the first iteration and $ p_{ref}(\bm{x})=h(\bm{x}, \widehat{\bm{v}}^{(t-1)})$ thereafter.
(2) Calculate the LSF value $g(\cdot)$ for each $\bm{x}_k$. Set $\gamma^{t}$ as the sample $\rho$-quantile of $\{g(\bm{x}_k),k=1,...,N\}$. $\rho$ represents a hyperparameter of the CE method and is typically chosen between 0.01 and 0.1 \cite{Kroese&others2013}.
(3) Solve the optimization problem of Eq.(\ref{Eq: samp. counterpart for intermediate target dist. in CE}) with $\mathcal{P}^{(t)}$ to get a new parameter vector $\widehat{\bm{v}}^{(t)}$. The above three steps are iterated until for some iteration $T$, $\gamma^{(T)}\leq0$. One then sets $\gamma^{(T)}=0$ and carries out step (3) one last time to get $\widehat{\bm{v}}^{(T)}$.\\      
In the iCE method, the intermediate target distributions are defined as:  
\begin{equation}
\label{Eq: intermediate target distribution of iCE method}
p^{(t)}(\bm{x}) \triangleq \frac{1}{Z^{(t)}} p_{\bm{X}}(\bm{x})\Phi \left( -\frac{g(\bm{x})}{\sigma^{(t)}} \right), t=1,...,T
\end{equation}    
where $\sigma^{(t)}>0$ and $\Phi$ is the cumulative distribution function (CDF) of the standard normal distribution. Note that $\lim\limits_{\sigma \rightarrow 0}(\Phi(-\frac{g(\bm{x})}{\sigma}))=\mathbb{I}\{g(\bm{x})\leq0\}$, meaning that for a decreasing sequence $\sigma^{(1)}> \cdots > \sigma^{(T)}$, the sequence of distributions gradually approaches the optimal IS distribution $p^*_{\bm{X}}(\bm{x})$. We note that alternative smooth approximations of the indicator function could be used instead of $\Phi$ to define the intermediate target distributions \cite{Uribe&others2021}.\\
The CE optimization problem for Eq.(\ref{Eq: intermediate target distribution of iCE method}) reads
\begin{equation}
\label{Eq: CE optimization for intermediate target dist. in iCE}
\bm{v}^{(t,*)} = \argmax\limits_{\bm{v}\in\mathcal{V}} 
\mathbb{E}_p[ \Phi(-g(\bm{X})/\sigma^{(t)}) \ln(h(\bm{X}; \bm{v})) ].
\end{equation}
The sample counterpart of Eq.(\ref{Eq: CE optimization for intermediate target dist. in iCE}) can then be expressed as 
\begin{equation}
\label{Eq: samp. counterpart for intermediate target dist. in iCE}
\widehat{\bm{v}}^{(t)} = \argmax\limits_{\bm{v}\in\mathcal{V}} \frac{1}{N} \sum \limits_{k=1}^{N} \frac{ p_{\bm{X}}(\bm{x}_k)\Phi(-g(\bm{x}_k)/\sigma^{(t)}) }{p_{ref}(\bm{x}_k)}\ln(h(\bm{x}_k; \bm{v})), \quad \bm{x}_k \sim p_{ref}(\cdot).
\end{equation}
According to \ref{Appendix: Self-normalized importance sampling and cross entropy method}, when $h(\bm{x};\bm{v})$ represents a properly parameterized exponential family, $\widehat{\bm{v}}^{(t)}$ is a self-normalized IS estimator of $\bm{v}^{(t,*)}$, independent of the choice of the intermediate target distributions. For the iCE method, the weight function of the self-normalized IS estimator of $\bm{v}^{(t,*)}$ equals
\begin{equation}
\label{Eq: the expression of weights}
W(\bm{x}; \sigma^{(t)})=\frac{p_{\bm{X}}(\bm{x})\Phi(-g(\bm{x})/\sigma^{(t)})}{p_{ref}(\bm{x})}.
\end{equation}
A common choice for measuring the accuracy of a self-normalized IS estimator is the ESS, whose approximate expression is given in Eq.(\ref{Eq: ESS}). With predefined sample size $N$, ESS is only a function of the c.o.v. of the weight, $\delta\left( W(\bm{X}; \sigma^{(t)})\right), \bm{X}\sim p_{ref}(\bm{x})$, which further depends on the reference distribution $p_{ref}(\bm{x})$ and the parameter $\sigma^{(t)}$.\\
In the $t$-th iteration of iCE, one fixes the reference distribution $p_{ref}(\bm{x})$ as $h(\bm{x}; \widehat{\bm{v}}^{(t-1)})$ (as $p_{\bm{X}}(\bm{x})$ in the first iteration) and selects $\sigma^{(t)}$ such that the sample c.o.v. of the weights $\{W(\bm{x}_k; \sigma^{(t)})\}_{k=1}^{N}$ equals a predefined target value $\delta_{tar}$, i.e., one solves the following optimization problem
\begin{equation}
\label{Eq: the updating rule of sigma}
\sigma^{(t)} = \argmin \limits _{\sigma \in(0, \sigma^{(t-1)})} |\widehat{\delta} \left( \{ W(\bm{x}_k;\sigma) \}_{k=1}^{N} \right) -\delta_{tar}|, \quad\quad \bm{x}_k \sim p_{ref}(\bm{x}).
\end{equation}
where $W(\bm{x}_k;\sigma)$ represents the weight in Eq.(\ref{Eq: the expression of weights}) and is a function of the optimization variable $\sigma$. In this way, the sample ESS equals $\frac{N}{1+\delta^2_{tar}}$. Hence, the accuracy of the self -normalized IS estimator $\widehat{\bm{v}}^{(t)}$ is tuned by the hyperparameter $\delta_{tar}$. A large $\delta_{tar}$ leads to an inaccurate $\widehat{v}^{(t)}$, while a small $\delta_{tar}$ increases the number of the intermediate target distributions $p^{(t)}(\bm{x})$ required to approach the optimal IS distribution $p^*_{\bm{X}}(\bm{x})$, thereby reducing the overall efficiency of the iCE method. This will be illustrated in detail in Section \ref{sec: Categorical parametric model}. In general, 1.5 is a good choice for $\delta_{tar}$ in the iCE method \cite{Papaioannou&others2019}. Once $\sigma^{(t)}$ is fixed, the optimization problem of Eq.(\ref{Eq: samp. counterpart for intermediate target dist. in iCE}) can be solved for the parameter vector $\widehat{\bm{v}}^{(t)}$. The corresponding distribution $h(\bm{x};\widehat{\bm{v}}^{(t)})$ is then used as the reference distribution for the $(t+1)$-th iteration.\\
The above procedure is iterated until the c.o.v. of the likelihood ratio \cite{Owen2013} for sampling from $p^{(t)}(\bm{x})$ instead of $p^*_{\bm{X}}(\bm{x})$ is smaller than $\delta_{\epsilon}$, i.e., 
\begin{equation}
\label{Eq: stopping criterion of iCE}
\delta \left( \frac{p^*_{\bm{X}}(\bm{X})}{p^{(t)}(\bm{X})} \right)=\delta \left( \frac{\mathbb{I}\{g(\bm{X})\leq0\}}{\Phi(-g(\bm{X})/\sigma^{(t)})} \right) \leq \delta_{\epsilon}, \quad\quad\bm{X}\sim p^{(t)}(\bm{x}). 
\end{equation}
In practice, we sample $\mathcal{P}^{(t)}=\{\bm{x}_k\}_{k=1}^{N}$ from $h(\bm{x};\widehat{\bm{v}}^{(t)})$ rather than $p^{(t)}(\bm{x})$, and check whether the sample c.o.v. of $\frac{\mathbb{I}\{g(\bm{x}_k)\leq0\}}{\Phi(-g(\bm{x}_k)/\sigma^{(t)})}$ is less or equal than $\delta_{\epsilon}$. Typically, $\delta_{\epsilon}$ is chosen the same as $\delta_{tar}$ \cite{Papaioannou&others2019}.\\
The algorithm for the iCE method is shown in Algorithm \ref{Alg: iCE}.    
\begin{algorithm}[!htbp]
\caption{Improved cross entropy algorithm} \label{Alg: iCE}
\LinesNumbered
\KwIn{$N$, $\delta_{tar}$, $\delta_{\epsilon}$}

$t \gets 1$, $t_{max} \gets 50$, $\sigma_0 \gets \infty$\\
$h(\bm{x}; \widehat{\bm{v}}^{(t-1)}) \gets p_{\bm{X}}(\bm{x})$\\ 
\While {true}{ 
	Generate $N$ samples $\{ \bm{x}_k \}_{k=1}^{N}$ from $h(\bm{x}; \widehat{\bm{v}}^{(t-1)})$ and calculate the corresponding LSF values $\{g(\bm{x}_k)\}_{k=1}^{N}$\\
	Compute the sample c.o.v. $\widehat{\delta}$ of $\left\{ \frac{\mathbb{I}\{ g(\bm{x}_k)\leq0 \}}{\Phi(-g(\bm{x}_k)/\sigma^{(t-1)})} \right\}_{k=1}^{N}$\\	 
	\If{ $t>t_{max} $ or $ \widehat{\delta} \leq \delta_{\epsilon}$}{Break
	}
	Determine $\sigma^{(t)}$ through solving Eq.(\ref{Eq: the updating rule of sigma})\\
	Compute $\widehat{\bm{v}}^{(t)}$ through solving Eq.(\ref{Eq: samp. counterpart for intermediate target dist. in iCE})\\
	$t \gets t+1$
}
$T \gets t-1$\\
Use $h(\bm{x}; \widehat{\bm{v}}^{(T)})$ as the IS distribution and calculate the IS estimator $\widehat{p}_f$ through Eq.(\ref{Eq: IS estimator})\\
\KwOut{$\widehat{p}_f$}
\end{algorithm}
\section{Bayesian cross entropy method for the categorical parametric family}
\label{sec: Categorical parametric model}
In this section, we consider the iCE method for estimating a rare event with a discrete random input $\bm{X}$, which often occurs in network reliability assessment. For discrete inputs $\bm{X}$, the probability mass function of $\bm{X}$, $p_{\bm{X}}(\bm{x})$, defines the probability of the corresponding outcome, i.e., $p_{\bm{X}}(\bm{x}) = \Pr(\bm{X} = \bm{x})$. We consider a slightly different definition of the intermediate target distribution in Eq.(\ref{Eq: intermediate target distribution of iCE method}), which results from the definition of an auxiliary LSF $g_a(\bm{x})$:
\begin{equation}
\label{Eq: auxiliary LSF}
g_{a}(\bm{x}) \triangleq 
\begin{cases}
g(\bm{x}), & \text{if} \quad g(\bm{x})>0 \\
0,    & \text{if} \quad g(\bm{x})\leqslant 0
\end{cases}.
\end{equation}
Note that the failure probability $p_f$ is unchanged if the original LSF in Eq.(\ref{Eq: Failure probability}) is substituted with the auxiliary one, so we can equivalently estimate the probability that $g_{a}(\bm{X})\leq0$ for $p_f$, i.e., 
\begin{equation}
\label{Eq: failure probability of the auxiliary LSF}
p_f = \mathbb{P}(g_a(\bm{X})\leq0) = \sum_{\bm{x} \in \Omega_{\bm{X}}} p_{\bm{X}}(\bm{x}) \mathbb{I}\{ g_a(\bm{x})\leq0 \},
\end{equation}
where $\Omega_{\bm{X}}$ is the sample space of the input random variables. In this way, the intermediate target distribution in Eq.(\ref{Eq: intermediate target distribution of iCE method}) becomes
\begin{equation}
\label{Eq: modified intermediate target distributions}
p^{(t)}(\bm{x}) \triangleq \frac{1}{Z^{(t)}} p_{\bm{X}}(\bm{x})\Phi \left( -\frac{g_a(\bm{x})}{\sigma^{(t)}} \right), t=1,...,T.
\end{equation}
In the following, we discuss the properties of the iCE method with the intermediate target distribution in Eq.(\ref{Eq: modified intermediate target distributions}). 
In particular, we examine the adaptation of the intermediate target distribution following Eq.(\ref{Eq: the updating rule of sigma}) and formulate a theorem stating that, under certain assumptions, the resulting distribution sequence gradually approaches the optimal IS. For the independent categorical parametric family, i.e., the joint distribution consisting of independent components that follow the categorical distribution, we further illustrate the overfitting issue of the standard iCE method. Based on this observation, we introduce a novel approach called the Bayesian improved cross entropy (BiCE) method that circumvents this problem. 
\subsection{Adaptation of the intermediate target distribution}
\label{subsec: Adaptation of the intermediate target distribution}
\begin{figure}[!h]
    \centering
	\includegraphics[scale=0.6]{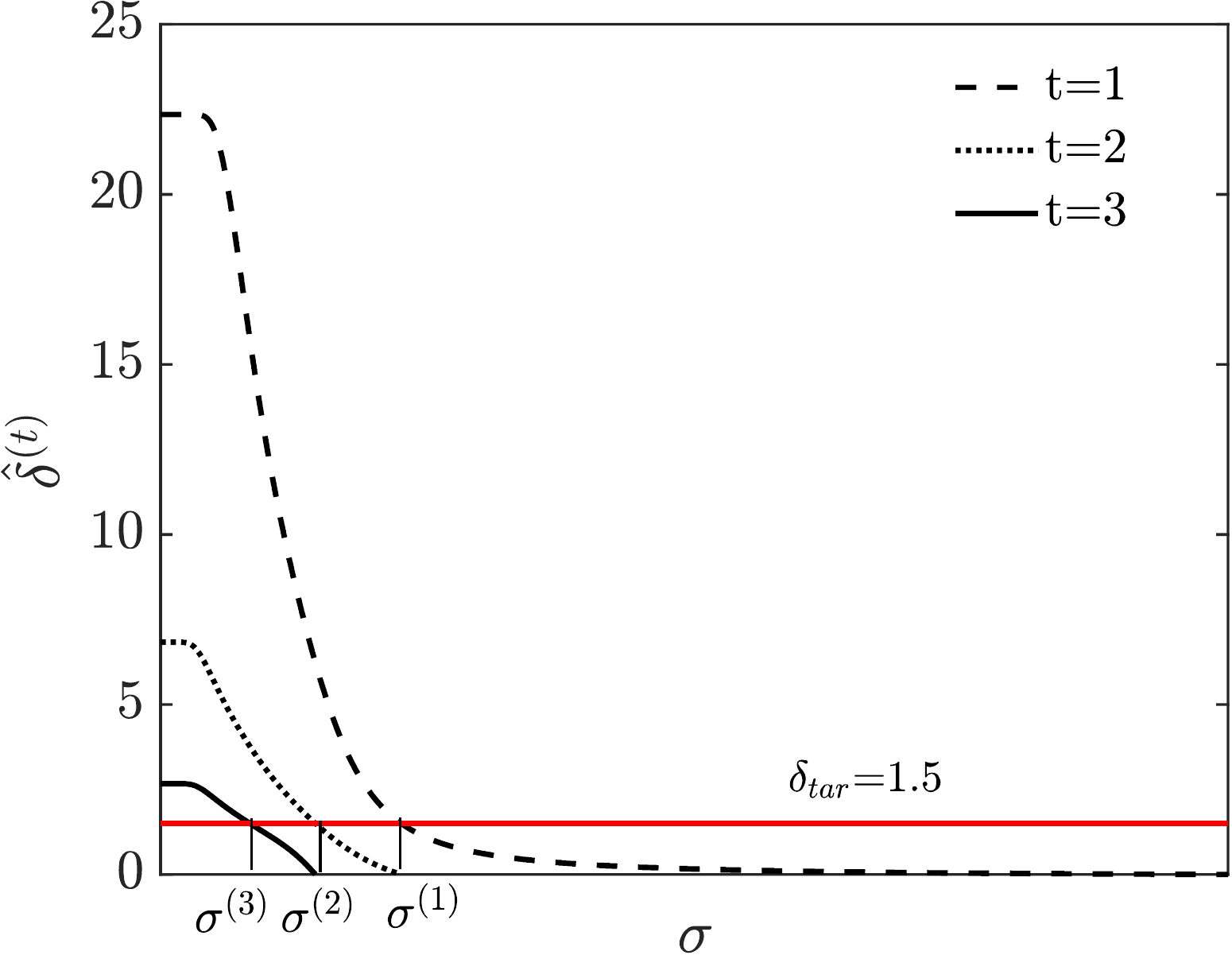}
    \caption{Schematic diagram of the adaptation of the $\sigma^{(t)}$.}
	\label{Fig: Adaptation of the intermediate distribution}
\end{figure}

The adaptation of the intermediate target distribution $p^{(t)}(\bm{x})$, or equivalently the parameter $\sigma^{(t)}$, plays an important role in achieving a balance between the efficiency and accuracy of the iCE method. Therefore, it is worth taking a closer look at the updating formula of $\sigma^{(t)}$ in Eq.(\ref{Eq: the updating rule of sigma}) and (\ref{Eq: stopping criterion of iCE}). To simplify the problem, we make the following assumptions:
\begin{assum}
The intermediate target distributions $p^{(t)}(\bm{x}) ,t=1,...,T$ are included in the parametric family $ h(\bm{x}; \bm{v})$ and therefore can be perfectly matched by $h(\bm{x}; \bm{v}^{(t,*)}), t=1,...,T$.
\end{assum}
\begin{assum}
The sample size is infinite such that $\widehat{\bm{v}}^{(t)}$ is the same as $\bm{v}^{(t,*)}$.
\end{assum}
Under these two assumptions, the sample c.o.v. of the weight in Eq.(\ref{Eq: the updating rule of sigma}) converges to the true c.o.v. $\delta \left( \frac{\Phi(-g_a(\bm{X})/\sigma)}{\Phi(-g(\bm{X})/\sigma^{(t-1)})} \right),\bm{X}\sim p^{(t-1)}(\bm{x})$, which is a function of $\sigma$. We write this function as $\delta^{(t)}(\sigma)$ for the rest of this paper. According to Eq.(\ref{Eq: stopping criterion of iCE}), the adaptive procedure of iCE method is stopped when $\delta^{(t)}(0)\leq \delta_{\epsilon}$. In \cite{Chan&others2022b}, we introduce the following two theorems
\begin{theor}
Under Assumptions 1 and 2, $\delta^{(t)}(\sigma)$ is a strictly decreasing function of $\sigma$ over $[0, \sigma^{(t-1)}]$. 
\end{theor}
\begin{theor}
Under Assumptions 1 and 2, it holds $\delta^{(t)}(\sigma^{(t-1)})=0$ and $\delta^{(t)}(0)=\sqrt{\frac{Z^{(t-1)}}{0.5p_f}-1}>0$, where $Z^{(t-1)}$ is the normalizing constant of $p^{(t-1)}(\bm{x})$ and can be expressed as 
\begin{equation}
\label{Eq: the expression of Z}
Z^{(t-1)} = 0.5p_f+\sum_{g_{a}(\bm{x})>0} p_{\bm{X}}(\bm{x})\Phi \left( -\frac{g_{a}(\bm{x})}{\sigma^{(t-1)}} \right). 
\end{equation}
\end{theor}
As a corollary, the optimization problem of Eq.(\ref{Eq: the updating rule of sigma}) has a unique solution $\sigma^{(t)}$ that is smaller than $\sigma^{(t-1)}$, resulting in a decreasing sequence of $\sigma$, i.e., $\sigma^{(1)}> \sigma^{(2)}>\cdots>\sigma^{(t)}$. Additionally, since $Z^{(t)}$ is a strictly increasing function of $\sigma^{(t)}$ according to Eq.(\ref{Eq: the expression of Z}), we have $Z^{(1)}> Z^{(2)} >\cdots> Z^{(t)}$, and this further leads to another decreasing sequence of $\delta^{(t)}(0)$, i.e., $\delta^{(1)}(0)>\delta^{(2)}(0)>\cdots>\delta^{(t)}(0)$. If for some iteration $T$ it holds $\delta^{(T)}(0) \leq \delta_{\epsilon}$, we terminate the adaptive procedure of iCE. The adaptation of $\sigma^{(t)}$ under Assumptions 1 and 2 is intuitively illustrated in Fig. \ref{Fig: Adaptation of the intermediate distribution}. All symbols in the figure have the same meaning as before.\\
In practice, the parametric family $h(\bm{x}; \bm{v})$ has limited flexibility, and the sample size is also finite due to limited computational budgets. Nevertheless, we expect that the results given in Theorems 4.1 and 4.2 still apply when $h(\bm{x}; \bm{v}^{(t)})$ forms a good approximation of $p^{(t)}(\bm{x})$, which is confirmed in the numerical experiments in Section \ref{sec: Examples}. 
The adaptation of $p^{(t)}(\bm{x})$ in iCE is tuned by hyper-parameters $\delta_{tar}$ and $\delta_{\epsilon}$. A small $\delta_{\epsilon}$ indicates a strict convergence criterion for $p^{(t)}(\bm{x})$, which leads to a more accurate yet less efficient result. 
It should be stressed that if the capacity of the parametric family is insufficient, the algorithm often fails to converge with a small $\delta_{\epsilon}$. Similarly to $\delta_{\epsilon}$, a small $\delta_{tar}$ leads to a $\sigma^{(t)}$ close to $\sigma^{(t-1)}$, which lowers the speed of the intermediate target distribution approaching the optimal IS distribution, thereby reducing the overall efficiency of the iCE algorithm. Conversely, a small $\delta_{tar}$ implies a large ESS and hence high accuracy of $\widehat{\bm{v}}^{(t)}$ in each $t$-th iteration of iCE method. 
In this paper, we suggest select $\delta_{tar}=\delta_{\epsilon}$ from 1 to 2, which is justified by the numerical examples in Sec. \ref{sec: Examples}. 
We also find that, instead of the original weight function defined in Eq.(\ref{Eq: the expression of weights}), using the alternative weight function 
\begin{equation}
\label{Eq: weight function}
W^{alt}(\bm{x}; \sigma) \triangleq \frac{\Phi(-g_a(\bm{x})/\sigma)}{\Phi(-g_a(\bm{x})/\sigma^{(t-1)} )}
\end{equation}
when solving $\sigma^{(t)}$ through Eq.(\ref{Eq: the updating rule of sigma}) will lead to a better convergence of the iCE algorithm for network reliability assessment, especially when $\delta_{\epsilon}$ is small. 

\subsection{Parametric distribution family for discrete inputs}
\label{subsec: Zero count problem}
To form a good approximation of $p^{(t)}(\bm{x})$ in the iCE method, a proper choice of the parametric family is necessary. In the context of reliability of systems with multi-state components, the obvious choice of the parametric model is the multivariate categorical distribution, which assigns a probability to each system state of the network, i.e., to each possible state in the sample space of the input distribution. The multivariate categorical distribution has great flexibility as it includes all possible distributions defined in the sample space of the network components. However, the number of parameters of this model grows exponentially with the input dimension (number of components) making this model impractical even for moderate dimensions. Therefore, consider independent categorical distributions in the following.\\ 
Suppose $\bm{X}$ is a $n$ dimensional input random vector with statistically independent components and each $d$-th component $X_d$ follows the categorical distribution taking values $\{s_{d,1}, \cdots s_{d,n_d} \}$ with probabilities $\{p_{d,1},\cdots,p_{d,n_d} \}$. $n_d$ is the number of sample states of $X_d$. The independent categorical family for $\bm{X}$ has the following general form:
\begin{equation}
\label{Eq: Independent categorical parametric family}
h(\bm{x}; \bm{v}) = \prod\limits_{d=1}^n h_d(x_d; \bm{v}_d)= \prod\limits_{d=1}^n \prod\limits_{i=1}^{n_d}  
v_{d,i}^{\mathbb{I}\{ x_d = s_{d,i} \}}, 0 \leq v_{d,i} \leq 1, \sum_{i=1}^{n_d} v_{d,i}=1,
\end{equation}
where $h_d(x_d; \bm{v}_d)=\prod\limits_{i=1}^{n_d}  
v_{d,i}^{\mathbb{I}\{ x_d = s_{d,i} \}}$ represents a univariate categorical distribution for $X_d$ that assigns a probability of $v_{d,i}$ to each $i$-th state $s_{d,i}$ of $X_d$, and $\bm{v} = \{ \bm{v}_d\}_{d=1}^{n}$ gathers the parameters of all components of the independent categorical family $h(\bm{x}; \bm{v})$.\\
The optimal parameter $\bm{v}^{(t,*)}$ is obtained through solving Eq.(\ref{Eq: CE optimization for intermediate target dist. in iCE}), which gives\cite{Rubinstein&Kroese2016}
\begin{equation}
\label{Eq: v_star_categorical}
v^{(t,*)}_{d,i}=\mathbb{E}_{p^{(t)}}[\mathbb{I}\{X_d=s_{d,i} \}].
\end{equation}
The explicit expression of $\bm{v}^{(t,*)}$ requires knowledge of the normalizing constant of $p^{(t)}$ and hence cannot be directly used in iCE method. Through optimizing the alternative objective function in Eq.(\ref{Eq: samp. counterpart for intermediate target dist. in iCE}), the near-optimal parameter $\widehat{\bm{v}}^{(t)}$ is explicitly given by 
\begin{equation}
\label{Eq: v_hat_categorical}
\widehat{v}^{(t)}_{d,i}= \frac{ \sum_{k=1}^{N} W(\bm{x}_k; \sigma^{(t)})\mathbb{I}\{ x_{k,d} = s_{d,i} \} }{\sum_{k=1}^{N} W(\bm{x}_k; \sigma^{(t)})}, \quad d=1,...,n, \quad i=1,...,n_d,
\end{equation}
where samples $\{\bm{x}_k \}_{k=1}^{N}$ are generated from the reference distribution $p_{ref}(\bm{x})=h(\bm{x}; \widehat{\bm{v}}^{(t-1)})$, and $W(\bm{x}_k; \sigma^{(t)})= \frac{p_{\bm{X}}(\bm{x}_k) \Phi(-g_a(\bm{x_k})/\sigma_t) }{p_{ref}(\bm{x}_k)}$. The expression of Eqs.(\ref{Eq: v_star_categorical}) and (\ref{Eq: v_hat_categorical}) can also be obtained by considering that the independent categorical distribution is a member of the exponential family \cite{Murphy2012}.
Note that $\widehat{\bm{v}}^{(t)}$ is the self-normalized estimator of $\bm{v}^{(t,*)}$. Additionally, $\widehat{\bm{v}}^{(t)}$ can be regarded as the weighted MLE of $\bm{v}$. This is because the objective function in Eq.(\ref{Eq: samp. counterpart for intermediate target dist. in iCE}) can be interpreted as a weighted log-likelihood function $\mathcal{LL}(\bm{v})$, with data set $\{ \bm{x}_k \}_{k=1}^{N}$ and weights $\{W(\bm{x}_k)/N\}_{k=1}^{N}$.\\ 
Similarly to the MLE of an independent categorical distribution, $\widehat{\bm{v}}^{(t)}$ suffers from overfitting, which is also known as the zero count problem in the context of MLE with categorical data \cite{Murphy2012}, and results in poor performance when the sample size is small. 
In particular, if there is no sample whose $d$-th component equals $s_{d,i}$, $s_{d,i}$ will be assigned a zero probability according to Eq.(\ref{Eq: v_hat_categorical}), i.e., $\hat{v}^{(t)}_{d,i} = 0$. In the context of the iCE method, the parameter vector $\widehat{\bm{v}}^{(t)}$ is employed to generate samples at the $(t+1)$ iteration, and hence, $s_{d,i}$ will not occur in any of the new generated samples. In this way, we have $\hat{v}^{(t)}_{d,i}=\hat{v}^{(t+1)}_{d,i}=\cdots =\hat{v}^{(T)}_{d,i}=0$, resulting in a reduced sample space of the final IS distribution $h(\bm{x}; \widehat{\bm{v}}^{(T)})$. However, for the optimal IS distribution, state $s_{d,i}$ is not necessarily negligible. In other words, the reduced sample space may only cover a part of the failure domain $F$, thereby underestimating the failure probability $p_f$.
\subsection{Bayesian improved cross entropy method}
\label{subsec: Bayesian improved cross entropy method}
In this subsection, we propose an accurate yet efficient algorithm termed Bayesian improved cross entropy method (BiCE) that circumvents the zero count problem. In this approach, instead of employing a weighted MLE estimator $\widehat{\bm{v}}^{(t)}$, a prior distribution is imposed on $\bm{v}^{(t)}$, and the posterior predictive distribution is derived, which is then employed to update the independent categorical family in iCE.\\ 
We insert the expression of the independent categorical parametric family of Eq.(\ref{Eq: Independent categorical parametric family}) into Eq.(\ref{Eq: samp. counterpart for intermediate target dist. in iCE}) and rewrite the objective function, or the weighted log-likelihood function $\mathcal{LL}(\bm{v})$, as follows
\begin{align}
\label{Eq: decomposition of the log-likelihood function}
\mathcal{LL}(\bm{v}) 
&= \sum_{k=1}^{N} \frac{W(\bm{x}_k; \sigma^{(t)})}{N} \ln \left( \prod_{d=1}^{n}h_d(x_{k,d}; \bm{v}_d) \right)  \notag\\
&= \sum_{d=1}^n\sum_{k=1}^{N} \frac{W(\bm{x}_k; \sigma^{(t)})}{N} \ln \left( h_d(x_{k,d}; \bm{v}_d) \right) \notag\\
& \triangleq \sum_{d=1}^n \mathcal{LL}_d(\bm{v}_d),
\end{align}
where $\mathcal{LL}_d(\bm{v}_d)$ is the weighted log-likelihood function of a one dimensional categorical family $h_d(x_d; \bm{v}_d)$, with data set $\{x_{k,d} \}_{k=1}^{N}$ and weights $\{ \frac{ W(\bm{x}_k;\sigma^{(t)}) }{N} \}_{k=1}^{N}$. From Eq.(\ref{Eq: decomposition of the log-likelihood function}), we find that, once the sample set is fixed, the parameter vectors $\bm{v}_d, d=1,\cdots,n$, are decoupled from each other in the expression of $\mathcal{LL}(\bm{v})$, that is the influence of each $\bm{v}_d$ on the outcome of $\mathcal{LL}(\bm{v})$ is separated (or additive). Additionally, we note that the feasible region for each parameter vector $\bm{v}_d$, that is $\mathcal{V}_d:\{ 0 \leq v_{d,i} \leq 1; i=1,\cdots,n_d | \sum_{i=1}^{n_d} v_{d,i}=1 \}$, is independent. Therefore, the original optimization problem can be decomposed into $n$ simpler subproblems, in which $\mathcal{LL}_d(\bm{v}_d)$ is maximized with respective to $\bm{v}_d \in \mathcal{V}_d$. The solutions to the subproblems are then concatenated to give a solution to the original problem, i.e., $\widehat{\bm{v}}^{(t)}= [\widehat{\bm{v}}^{(t)}_1; \cdots; \widehat{\bm{v}}^{(t)}_n]$. Therefore, it is sufficient to discuss the following subproblem:
\begin{align}
\widehat{\bm{v}}_d^{(t)}
&= \argmax\limits_{\bm{v}_d \in \mathcal{V}_d} \sum_{k=1}^{N} \frac{W(\bm{x}_k; \sigma^{(t)})}{N}  \ln \left( h_d(x_{k,d}; \bm{v}_d) \right) \notag. 
\end{align}
Note that multiplying the objective function with a positive constant $\alpha$ or taking an exponential of the objective function does not change the solution to the optimization problem, so we have 
\begin{align}
\widehat{\bm{v}}_d^{(t)}
&= \argmax\limits_{\bm{v}_d \in \mathcal{V}_d} \sum_{k=1}^{N} \frac{ \alpha W(\bm{x}_k; \sigma^{(t)})}{N}  \ln \left( h_d(x_{k,d}; \bm{v}_d) \right) \notag\\ 
\label{Eq: weighted likelihood function} 
&= \argmax\limits_{\bm{v}_d \in \mathcal{V}_d} \prod_{k=1}^{N}   \left( h_d(x_{k,d}; \bm{v}_d) \right)^{\frac{ \alpha W(\bm{x}_k; \sigma^{(t)})}{N}}. 
\end{align}
The objective function in Eq.(\ref{Eq: weighted likelihood function}) can be regarded as a weighted likelihood function for $h_d(x_d; \bm{v}_d)$ with data set $\{ x_{k,d} \}_{k=1}^{N} $ and weights $\{ \frac{\alpha W(\bm{x}_k; \sigma^{(t)})}{N}\}_{k=1}^{N} $. In this work, $\alpha$ is chosen such that the weighted likelihood function coincides with the standard likelihood function with unit weights when $W(\bm{x}_k), k=1,...,N$ are all equal, which gives 
\begin{equation}
\label{Eq: choice of the alpha}
\alpha=\frac{N^2}{\sum_{k=1}^{N}W(\bm{x}_k;  \sigma^{(t)})}.
\end{equation}
Inserting the above expression of $\alpha$ and substituting the parametric family $h_d(x_{k,d}; \bm{v}_d)$ with the expression in Eq.(\ref{Eq: Independent categorical parametric family}) into Eq.(\ref{Eq: weighted likelihood function}) gives
\begin{equation}
\label{Eq: weighted likelihood function2}
\widehat{\bm{v}}_d^{(t)}
= \argmax\limits_{\bm{v}_d \in \mathcal{V}_d}
\prod\limits_{i=1}^{n_d} v_{d,i}^{ \sum_{k=1}^{N} \left(\mathbb{I}\{ x_{k,d} = s_{d,i} \} w_k \right) } \triangleq \argmax\limits_{\bm{v}_d \in \mathcal{V}_d} \mathcal{L}_d(\bm{v}_d), 
\end{equation}
where $ w_k \triangleq N \frac{W(\bm{x}_k; \sigma^{(t)})}{\sum_{k=1}^{N}W(\bm{x}_k; \sigma^{(t)})} $ is the weight for the $k$-th sample $\bm{x}_k$, and $\mathcal{L}_d(\bm{v}_d)$ is the weighted likelihood function for the categorical distribution $h_d(x_d; \bm{v}_d)$ with data set $\{ x_{k,d} \}_{k=1}^{N}$ and weights $\{w_k\}_{k=1}^{N}$. 
Note that when $W(\bm{x}_k; \sigma^{(t)}), k=1,...,N$ are all equal, $\mathcal{L}_d(\bm{v}_d)$ degenerates into a standard likelihood function with unit weights, i.e., $w_k=1, k=1,...,N$. The analytical solution to the optimization problem in Eq.(\ref{Eq: weighted likelihood function2}) is obtained as
\begin{equation}
\label{Eq: v_hat for univarite categorical}
\widehat{v}_{d,i}^{(t)}= \frac{1}{N} \sum_{k=1}^{N} w_k \mathbb{I}\{ x_{k,d}= s_{d,i} \}, \quad i=1,...,n_d,
\end{equation} 
which coincides with the expression of $\widehat{\bm{v}}^{(t)}$ in Eq.(\ref{Eq: v_hat_categorical}). Note that $0 \leq\widehat{v}_{d,i}\leq 1$, and $\sum_{i=1}^{n_d}\widehat{v}_{d,i}^{(t)}=1$.\\
The basic idea of the proposed BiCE method is to employ a Bayesian approach for estimating the parameter vector $\bm{v}_d$ that aims at avoiding the overfitting problem through adding a regularization term in Eq.(\ref{Eq: weighted likelihood function}). In particular, the weighted likelihood function in the second line of Eq.(\ref{Eq: weighted likelihood function}) is multiplied by a prior distribution over the parameter vector $\bm{v}_d$. 
Instead of solving the regularized optimization problem, we derive the full posterior predictive distribution for the categorical distribution. 
The imposed prior distribution acts as an additional information set, and through Bayes' rule, we combine the two sources of information. If one source contains more information than the other, the posterior distribution will be pulled towards it; the relative 'strength' between the prior and the data is adjusted by the $\alpha$ and also the prior parameters.\\ 
The prior $f'(\bm{\bm{v}_d})$ is chosen to be a Dirichlet distribution in this paper, which is the conjugate prior for the parameter vector $\bm{v}_d$ of a categorical distribution. The PMF of the Dirichlet distribution can be expressed as follows 
\begin{equation}
\label{Eq: Dirichlet prior}
f'(\bm{\bm{v}_d})=\text{Dir}(\bm{v}_d;\bm{\theta}_d) = \frac{1}{B(\bm{v}_d)} \prod_{i=1}^{n_d} (v_{d, i})^{\theta_{d,i}-1}\mathbb{I}\{ \bm{v}_d \in \mathcal{V}_d\},
\end{equation}
where $\bm{\theta}_d=\{ \theta_{d,i} >0 \}_{i=1}^{n_d}$ represents the parameter vector of the prior, and $B(\bm{v}_d)$ is the normalizing constant  with $B(\cdot)$ being the multivariate Beta function.\\ 
Combining Eq.(\ref{Eq: Dirichlet prior}) with the weighted likelihood in Eq.(\ref{Eq: weighted likelihood function2}), the posterior distribution $ f''(\bm{\bm{v}_d})$ is also a Dirichlet distribution. In fact, according to Bayes' rule, $ f''(\bm{\bm{v}_d})$ can be expressed as 
\begin{align}
f''(\bm{v}_d) 
&\propto f'(\bm{v}_d)\mathcal{L}_d(\bm{v}_d) \notag \\
&\propto \prod_{i=1}^{n_d} (v_{d, i})^{ N\widehat{v}_{d,i}^{(t)}+\theta_{d,i}-1} \mathbb{I}\{ \bm{v}_d \in \mathcal{V}_d\} \notag \\
\label{Eq: Dirichlet posterior}
&= \text{Dir}(\bm{v}_d; N\widehat{\bm{v}}_d^{(t)}+\bm{\theta_d} ). 
\end{align}
We then utilize the full distribution of the posterior $f''(\bm{v}_d)$ for updating the categorical parametric family $h_d(x_d; \bm{v}_d)$. That is, we calculate the probability $\widetilde{\mu}^{(t)}_{d,i}$ that each state $s_{d,i}$ of $h_d(x_d; \bm{v}_d)$ occurs under the Dirichlet posterior distribution in Eq.(\ref{Eq: Dirichlet posterior}). Based on the total probability theorem, $\widetilde{\mu}^{(t)}_{d,i}$ can be calculated through    
\begin{equation}
\widetilde{\mu}^{(t)}_{d,i}
=  \int_{\mathcal{V}_d} h_d(s_{d,i};\bm{v}_d) f''(\bm{v}_d) d\bm{v}_d 
=  \int_{\mathcal{V}_d} v_{d,i} \text{Dir}(\bm{v}_d; N\widehat{\bm{v}}_d^{(t)}+\bm{\theta_d}) d\bm{v}_d 
= \mathbb{E}_{\text{Dir}}[V_{d,i}]. \notag
\end{equation}
Since the mean value of a Dirichlet distribution $\text{Dir}(\bm{v}_d; \bm{\theta}_d)$ is explicitly given by  $\frac{\theta_{d,i}}{\sum_{j=1}^{n_d}\theta_{d,j}}, i=1,...,n_d$, \cite{Murphy2012}, we further have
\begin{align}
\label{Eq: posterior predictive distribution}
\widetilde{\mu}^{(t)}_{d,i} 
&= \frac{N \widehat{v}^{(t)}_{d,i} + \theta_{d, i}}{ \sum_{j=1}^{n_d} \left(N \widehat{v}^{(t)}_{d,j} + \theta_{d, j}\right) } \notag \\
&= \frac{N \widehat{v}^{(t)}_{d,i} + \theta_{d, i}}{ N + \sum_{j=1}^{n_d} \theta_{d, j} } \notag \\
&= \lambda_d \widehat{v}^{(t)}_{d,i} + (1-\lambda_d) \frac{\theta_{d, i}}{\sum_{j=1}^{n_d} \theta_{d, j}},
\end{align}
where $\lambda_d \triangleq \frac{N}{ N + \sum_{j=1}^{n_d} \theta_{d, j} }$ denotes the combination factor. According to Eq.(\ref{Eq: posterior predictive distribution}), this estimator $\widetilde{\mu}^{(t)}_{d,i}$ can be written as a linear combination of the weighted MLE estimator $\widehat{v}_{d,i}^{(t)}$, which exploits the information of the weighted samples, and the prior estimator $\frac{\theta_{d,i}}{\sum_{i=1}^{n_d}(\theta_{d,i})}$, which can explore a different range of the sample space. The relative 'strength' of $\widehat{v}^{(t)}_{d,i}$ is indicated by the combination factor $\lambda_d$ and is tuned by $\sum_{i=1}^{n_d}\theta_{d,i}$. Apparently, the smaller the $\sum_{i=1}^{n_d}\theta_{d,i}$, the more dominant the $\widehat{v}^{(t)}_{d,i}$, and vice versa. We therefore favour a balanced prior with an appropriately large $\sum_{i=1}^{n_d}\theta_{d,i}$, that will not dominate but can still deviate the potentially overfitted weighted MLE estimator $\widehat{v}^{(t)}_{d,i}$. 
A further investigation of the prior distribution is left for the future work and in this paper, we simply employ a symmetric Dirichlet prior for each $\bm{v}_d$, and set 
\begin{equation}
\label{Eq: a symmetric Dirichlet prior}
\theta_{d,j} = b; \quad \quad \quad j = 1,...,n_d, d = 1,...,n,
\end{equation}
where $b$ is the hyper parameter. We suggest choose a moderate $b$, e.g., 5 or 10 when $N$ is around thousand.\\  
Additionally, $\widetilde{\mu}^{(t)}_{d,i}$ converges to $\widehat{v}_{d,i}^{(t)}$ as the sample size $N$ approaches infinity. Considering that $\widehat{v}_{d,i}^{(t)}$ is the normalized IS estimator of the optimal CE parameter $v_{d,i}^{(t,*)}$ in Eq.(\ref{Eq: v_star_categorical}), both $\widetilde{\mu}^{(t)}_{d,i}$ and $\widehat{v}_{d,i}^{(t)}$ will converge to $v_{d,i}^{(t,*)}$. 
Therefore, similar to $\widehat{v}_{d,i}^{(t)}$, the accuracy of $\widetilde{\mu}^{(t)}_{d,i}$ is guaranteed for a large sample size. 
Conversely, $\widetilde{\mu}^{(t)}_{d,i}$ is positive even for a small sample size (at least $\frac{b}{N+b \cdot n_d}$), and hence, it does not suffer from the zero count problem as in the case of weighted MLE.\\ 
In this way, $\widetilde{\bm{\mu}}^{(t)}_d = (\widetilde{\mu}^{(t)}_{d,1},..., \widetilde{\mu}^{(t)}_{d,n_d} )$ forms a new parameter vector for the one-dimensional categorical family $h_d(x_d; \bm{v}_d)$. $h_d(x_d; \widetilde{\bm{\mu}}^{(t)}_d)$ is also known as the posterior predictive distribution in Bayesian statistics, so we term $\widetilde{\bm{\mu}}^{(t)}_d$ the Bayesian estimator of $\bm{v}_d$.
After obtaining the Bayesian estimator $\widetilde{\bm{\mu}}^{(t)}_{d}$ for each dimension $d=1,...,n$ through Eq.(\ref{Eq: posterior predictive distribution}), we concatenate the results to get the Bayesian estimator $\widetilde{\bm{\mu}}^{(t)}$ for the independent categorical distribution $h(\bm{x};\bm{v})$, i.e., $\widetilde{\bm{\mu}}^{(t)}= [\widetilde{\bm{\mu}}^{(t)}_1;...,\widetilde{\bm{\mu}}^{(t)}_d]$. The posterior predictive distribution $h(\bm{x};\widetilde{\bm{\mu}}^{(t)})$ is then employed as the reference distribution $p_{ref}(\bm{x})$ for the $(t+1)$-th iteration in iCE. The resulting algorithm is termed the Bayesian improved cross entropy method (BiCE) and is given in Algorithm \ref{Alg: BiCE}.
\begin{algorithm}[!htbp]
\caption{Bayesian improved cross entropy algorithm} 
\label{Alg: BiCE}
\LinesNumbered
\KwIn{$N$, $\delta_{tar}$, $\delta_{\epsilon}$,  prior parameters $\{\bm{\theta}_1,...,\bm{\theta}_n\}$ }

$t \gets 1$, $t_{max} \gets 50$, $\sigma_0 \gets \infty$\\
$h(\bm{x}; \widetilde{\bm{\mu}}^{(t-1)}) \gets p_{\bm{X}}(\bm{x})$\\ 
\While {true}{ 
	Generate $N$ samples $\{ \bm{x}_k \}_{k=1}^{N}$ from $h(\bm{x}; \widetilde{\bm{\mu}}^{(t-1)})$ and calculate the corresponding LSF values $\{g_a(\bm{x}_k)\}_{k=1}^{N}$\\
	Compute the sample c.o.v. $\widehat{\delta}$ of $\left\{ \frac{\mathbb{I}\{ g_a(\bm{x}_k)\leq0 \}}{\Phi(-g_a(\bm{x}_k)/\sigma^{(t-1)})} \right\}_{k=1}^{N}$\\	 
	\If{ $t>t_{max} $ or $ \widehat{\delta} \leq \delta_{\epsilon}$}{Break
	}
	Determine $\sigma^{(t)}$ through solving Eq.(\ref{Eq: the updating rule of sigma}) using the alternative weight function defined in Eq.(\ref{Eq: weight function}).\\
	Compute $\widetilde{\mu}^{(t)}_{d,i}$ through Eq.(\ref{Eq: posterior predictive distribution}), for each $d$ and $i$\\
	$t \gets t+1$
}
$T \gets t-1$\\
Use $h(\bm{x}; \widehat{\bm{v}^{(T)}})$ as the IS distribution and calculate the IS estimator $\widehat{p}_f$ through Eq.(\ref{Eq: IS estimator})\\
\KwOut{$\widehat{p}_f$}
\end{algorithm}
\begin{remark}
Instead of using the full posterior distribution, one can also utilize the mode of the posterior distribution $\widetilde{\bm{v}}^{(t)}_d$ as a point estimate of $\bm{v}_d$, which is known as the maximum a posteriori (MAP) estimator in Bayesian statistics. 
By definition, the MAP estimator can be expressed as
\begin{equation}
\label{Eq: MAP estimator1}
\widetilde{\bm{v}}_d^{(t)} = \argmax \limits_{\bm{v}_d \in \mathcal{V}_d} f''(\bm{v}_d).
\end{equation}
Substituting the posterior $f''(\bm{v}_d)$ with the expression in Eq.(\ref{Eq: Dirichlet posterior}) and then solving the optimization problem in Eq.(\ref{Eq: MAP estimator1}) with a Lagrange multiplier gives us
\begin{align}
\widetilde{v}_{d,i}^{(t)} 
&= \frac{N\widehat{v}_{d,i}^{(t)}+\theta_{d,i}-1}{N\sum_{i=1}^{n_d} \widehat{v}_{d,i}^{(t)}+\sum_{i=1}^{n_d}(\theta_{d,i}-1)} \notag \\
\label{Eq: MAP estimator2}
&=\frac{N}{N+\sum_{i=1}^{n_d}(\theta_{d,i}-1)}\widehat{v}_{d,i}^{(t)}+\frac{\sum_{i=1}^{n_d}(\theta_{d,i}-1)}{N+\sum_{i=1}^{n_d}(\theta_{d,i}-1)} \frac{\theta_{d,i}-1}{\sum_{i=1}^{n_d}(\theta_{d,i}-1)}.
\end{align}
Comparing Eq.(\ref{Eq: MAP estimator2}) and Eq.(\ref{Eq: posterior predictive distribution}), we find that the MAP estimator $\widetilde{v}_{d,i}^{(t)}$ with prior parameter $\theta_{d,i}>1$ is the same as the Bayesian estimator $\widetilde{\mu}_{d,i}^{(t)}$ with prior parameter $\theta_{d,i}-1$. When $\theta_{d,i} = 1$, the MAP estimator $\widetilde{v}_{d,i}^{(t)} $ reduces to the weighted MLE $\widehat{v}_{d,i}^{(t) }$. 
\end{remark}  
\begin{remark}
If the symmetric Dirichlet prior described in Eq.(\ref{Eq: a symmetric Dirichlet prior}) is applied in the BiCE method, it holds that $\widetilde{\mu}^{(t)}_{d,i} = \frac{N}{N+n_d}\widehat{v}^{(t)}_{d,i}+\frac{b}{N+b \cdot n_d}$ according to Eq.(\ref{Eq: posterior predictive distribution}), which means the probability that $X_d$ equals $s_{d,i}$ in the posterior predictive distribution is at least $\frac{b}{N+b \cdot n_d}$. To avoid this probability being close to zero, the number of states for $X_d$, $n_d$, cannot be too large. Otherwise, the zero count problem can still occur in the BiCE method. 
\end{remark}   
\section{Examples}
\label{sec: Examples}
\subsection{Toy example: system with linear limit state functions}
In this example, we consider a LSF $g_1(\bm{x})$, that is a linear combination of 50 random variables. The coefficients of the first and the last 10 random variables are set to be 2 and 0, respectively, while for the remaining random variables, the coefficients are fixed at 1. The LSF reads    
\begin{equation}
g_1(\bm{x}) = \sum_{d=1}^{10}2 \cdot x_d + \sum_{d=11}^{40} 1 \cdot x_d + \sum_{d=41}^{50}0 \cdot x_d.
\end{equation}
\subsubsection{Binary input} 
\begin{figure}[!h]
    \centering
	\includegraphics[scale=0.7]{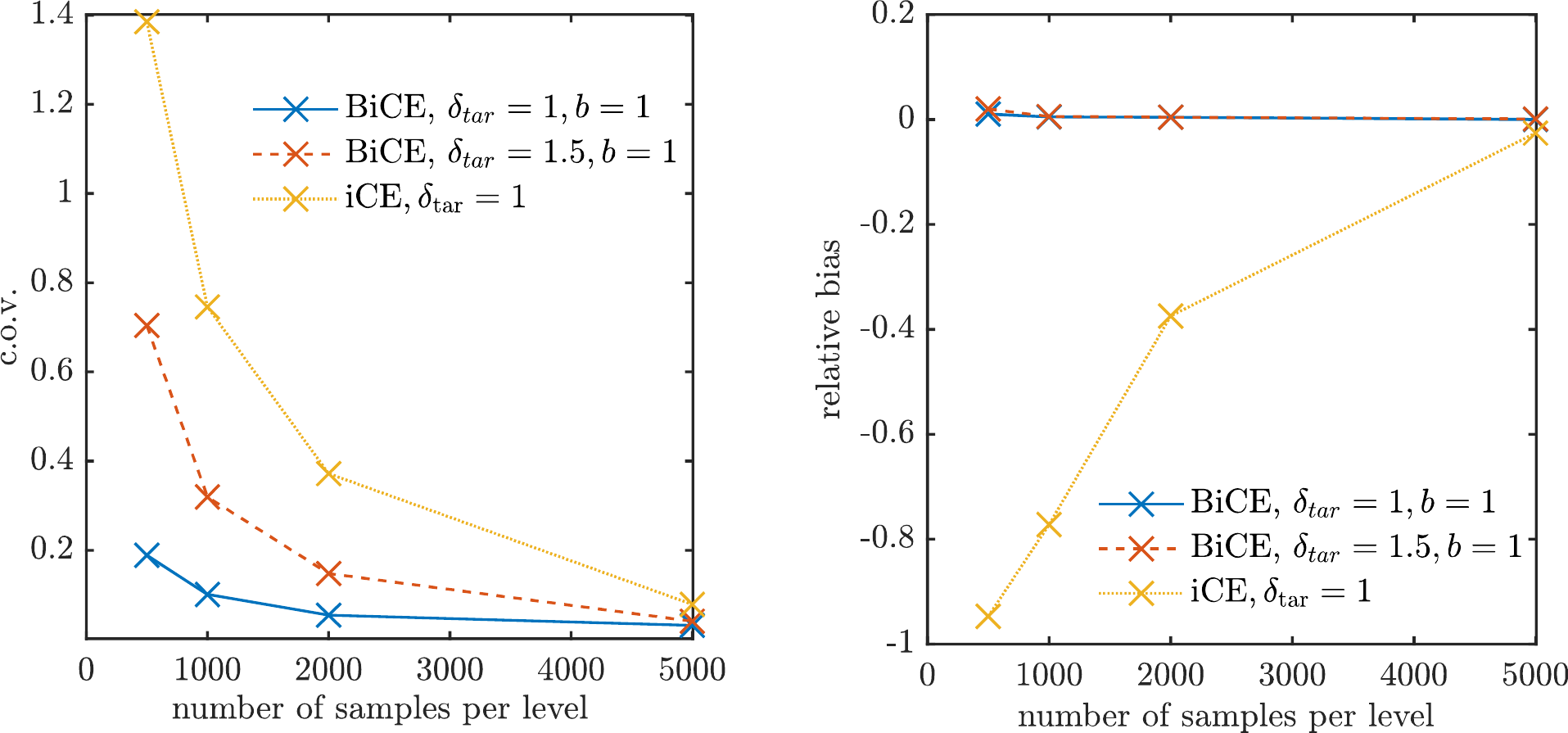}
    \caption{Parameter study for example 5.1.} 
	\label{Fig: Parameter study for example 5.1.}
\end{figure}

We first assume that $\{X_d\}_{d=1}^{50}$ are independent and identically distributed (i.i.d.) Bernoulli random variables with success probability $10^{-3}$, i.e., the probability that each $X_d$ takes the value 1 is $10^{-3}$. We estimate the probability that $g_1(\bm{X}) \geq 6$ using the BiCE and compare the result with that of the standard iCE approach. The exact solution is $1.387\cdot 10^{-7}$ through the convolution of two binomial distributions. We fix the sample size $N$ at $500$ and $2,000$ for BiCE and iCE, respectively, and set $\delta_{tar}=\delta_{\epsilon} = 1$ for both methods. $5,000$ repeated runs of each estimator are carried out to calculate the relative bias, the sample c.o.v., and the average computational cost (i.e., the average number of calls of the LSF) of the estimator. Also, the influence of different prior parameters on the performance of BiCE is investigated. We apply the symmetric Dirichlet prior defined in Eq.( \ref{Eq: a symmetric Dirichlet prior}), and vary the parameter $b$ therein. 
We note that when $n_d=2$ the Dirichlet distribution degenerates into the Beta distribution. The results are summarized in Table \ref{Table: performance of BiCE for Example 5.1.1}.\\
\begin{table}[!htbp]
    \centering
    \caption{Performance of BiCE for Example 5.1.1.}
    \label{Table: performance of BiCE for Example 5.1.1}
    \begin{tabular}{lllllll}
    \hline
        method & BiCE & BiCE & BiCE & BiCE & BiCE & iCE\\
        sample size, $N$ & $500$ & $500$ & $500$ & $500$ & $500$ & $2,000$\\
        prior parm., $b$ & $1$ & $5$ & $25$ &  $50$ & $500$ & $\slash$ \\
        \hline
        relative bias & 0.012 & 0.011 & 0.007 & 0.006 & -0.745 & -0.375\\
        sample c.o.v. & 0.196 & 0.122 & 0.287 & 0.787 &  21.209 & 0.372\\
        comp. cost & $4,660$ & $4,127$ & $2,495$ & $1,500$ & $1,700$ & $19,967$\\
        \hline  
    \end{tabular}
\end{table}
We can see from the table that the iCE method even with $N=2,000$ samples per level, which is four times larger than the number of samples used with the BiCE, performs poorly with a strong negative bias. This is due to the zero count problem described in Subsection \ref{subsec: Zero count problem}; the failure states of some of the input random variables are ignored throughout the sampling process, which leads to an under-representation of the failure domain. In contrast, BiCE with an uninformative prior, which is the case where $b=1$, works well, resulting in an efficient yet accurate estimator. The performance of the BiCE can be further enhanced though employing a larger $b$, e.g., the prior with $b=5$ outperforms the prior with $b=1$. Meanwhile, selecting an excessive value of $b$ leads to poor results.\\ 
To further illustrate the zero count problem, Fig. \ref{Fig: Parameter study for example 5.1.} shows the influence of the number of samples per level, $N$, on the c.o.v. and the relative bias of the BiCE and iCE estimates. In this figure, the blue solid line represents the BiCE method with uninformative prior and target c.o.v. $\delta_{tar}=\delta_{\epsilon}=1$; the red dashed line shows the BiCE method with uninformative prior and target c.o.v. $\delta_{tar}=\delta_{\epsilon}=1.5$. Both variants show a negligible relative bias for all considered $N$. In contrast, the relative bias of the standard iCE method is almost -100\% when the number of samples is small, i.e., $N=500$, and gradually approaches 0 as $N$ increases. Obviously, the zero count problem is less likely to happen for larger number of samples per intermediate level.
\subsubsection{Multi-state input}
\begin{figure}[!h]
    \centering
	\includegraphics[scale=0.68]{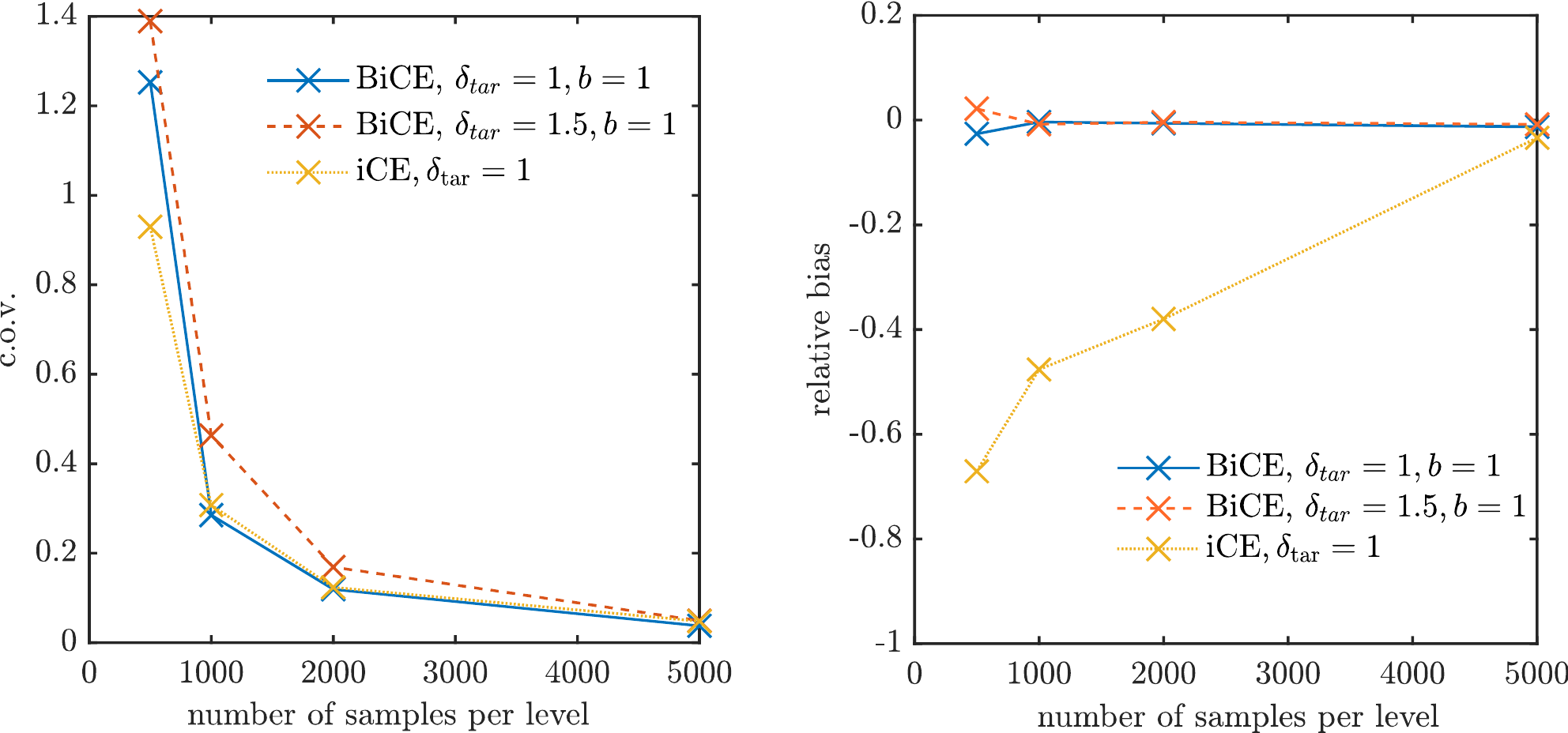}
    \caption{Parameter study for example 5.2.} 
	\label{Fig: Parameter study for example 5.2.}
\end{figure}

Next we assume that $\{X_d\}_{d=1}^{50}$ follows the i.i.d categorical distribution with categories 0, 1 and 3. The probabilities assigned to these categories are 0.899, 0.1, and $10^{-3}$. In this subsection, we estimate the probability that $g_1(\bm{x}) \geq 19$. The exact value of this probability is approximated through crude MCS with $10^7$ samples, resulting in $ 7.3 \cdot 10^{-5}$. 5000 repeated runs of BiCE with hyperparameters $\delta_{tar}=\delta_{\epsilon} =1$ and $N=1,000$ are performed. The obtained results are then compared with the standard iCE approach with $\delta_{tar}=\delta_{\epsilon} =1$ and $N=2,000$, and are summarized in Table \ref{Table: performance of BiCE for Example 5.1.2}. Similarly to the binary case in Subsection 5.1.1, the BiCE with the uniform prior, that is the case where $b=1$, outperforms the standard iCE approach, and the performance can be further improved through applying a larger $b$.\\
Fig. \ref{Fig: Parameter study for example 5.2.} shows the performance of the iCE and the BiCE methods for varying the number of samples per intermediate level for multi-state input. Similarly to Fig. \ref{Fig: Parameter study for example 5.1.}, which is given for binary input, the relative bias of the iCE method approaches zero as $N$ goes from 500 to $5,000$.
\begin{table}[!htbp]
    \centering
    \caption{Performance of BiCE for Example 5.1.2.}
    \label{Table: performance of BiCE for Example 5.1.2}
    \begin{tabular}{llll}
        \hline
        method & BiCE & BiCE & iCE\\
        sample size, $N$ & $1,000$ & $1,000$ & $2,000$\\
        prior parm., $b$ & $1$ & $10$ & $\slash$ \\
        \hline
        relative bias & -0.004 & -0.014 & -0.375 \\
        sample c.o.v. & 0.285 & 0.107 & 0.124 \\
        comp. cost & $7,484$ & $5,997$ & $15,966$ \\
        \hline  
    \end{tabular}
\end{table} 
\subsection{Multi-state two-terminal reliability}
\begin{figure}[!h]
    \centering
	\includegraphics[scale=0.5]{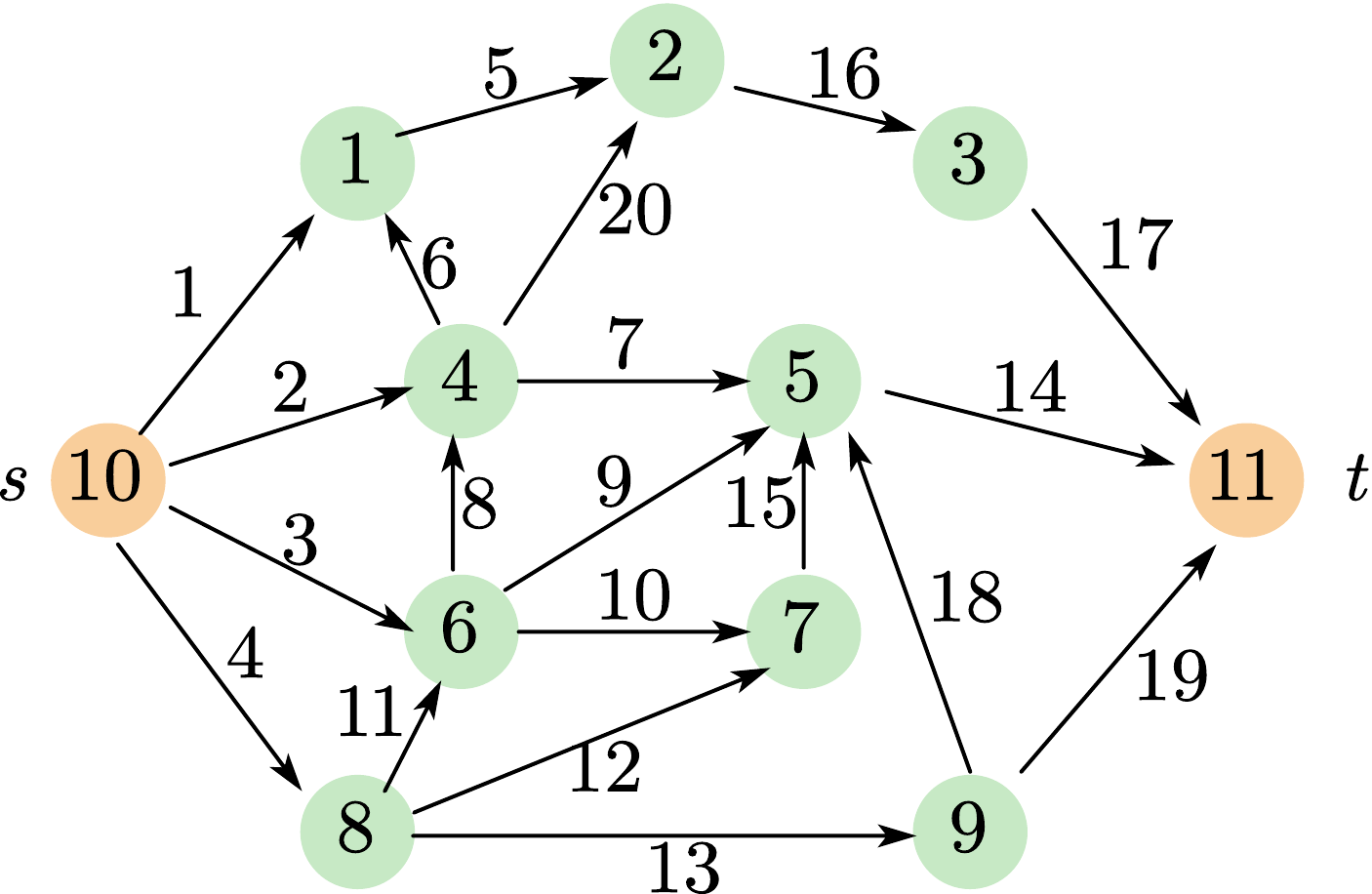}
    \caption{Topology of the network for example 5.2.}
	\label{Fig: topology of the network for example 5.2.}
\end{figure}

Fig.\ref{Fig: topology of the network for example 5.2.} shows the topology of a multi-state two-terminal network with 11 nodes and 20 edges (or arcs), which is motivated by the third example of \cite{Ramirez-marquez&Coit2005}. The capacity of each edge, that is, the maximum flow that can pass through the edge, takes the same three states 0, 3 and 5 with probability $10^{-3}$, 0.1, and 0.899, respectively. Also, the states of each edge are independent. We are required to estimate the probability that the maximum flow from the source node $s$ to the sink node $t$ is less or equal than a predefined demand $D_{tar} = 6$. The true value of the probability is approximately $1.8\cdot10^{-4}$ according to the crude MCS results (with sample size $2\cdot 10^6$), and this value is employed to assess the accuracy of the proposed BiCE approach. We choose $\delta_{tar}=\delta_{\epsilon} =1$ and $N=1,000$ and use an uninformative uniform prior, $\text{Dir}(\cdot; \bm{\theta}=[1, 1, 1]^T)$ for each dimension in BiCE, i.e., we set $b=1$. The relative bias, sample c.o.v., and average computational cost of 200 repeated runs of the BiCE are -0.0043, 0.100 and $9,385$, respectively, indicating an efficient yet accurate estimator. In contrast, the theoretical c.o.v. of the crude MCS with the same computational cost, i.e., with $9,385$ samples, equals 0.723, which is significantly larger than that of BiCE, and the standard iCE with hyperparameters $\delta_{tar}=\delta_{\epsilon} =1$ and $N=2,000$ will seriously underestimate the failure probability. 
On the other hand, the performance of the BiCE estimator can be improved through setting $b=10$. 
These results are summarized in Table \ref{Table: performance of BiCE for Example 5.2}. Table \ref{Table: the IS distribution averaged over 200 runs of BiCE for Example 5.2} shows the PMF of the final IS distribution in BiCE averaged over 200 repetitions. We can see from this table that the IS distribution of the 14-th edge differs the most from the input distribution in BiCE, followed by edges 16/17 and edges 4/13/19. This indicates that these are the most important edges for the failure probability. For remaining edges, the IS distribution differs only slightly from the input distribution.\\
\begin{table}[!htbp]
    \centering
    \caption{Performance of BiCE for Example 5.2.}
    \label{Table: performance of BiCE for Example 5.2}
    \begin{tabular}{lllll}
        \hline
        method & BiCE & BiCE & MCS & iCE\\
        sample size, $N$ & $1,000$ & $1,000$ & $9,385$ & $2,000$\\
        prior parm., $b$ & $1$ & $10$ & $\slash$ & $\slash$ \\
        \hline
        relative bias & -0.004 & -0.021 & 0 & -0.253 \\
        sample c.o.v. & 0.100 & 0.089 & 0.723 & 0.168 \\
        comp. cost & $9,385$ & $8,650$ & $9,385$ & $20,400$\\
        \hline  
    \end{tabular}
\end{table} 
\begin{table}[!htbp]
    \centering
    \caption{The PMF of the IS distribution in BiCE for Example 5.2 \\ (averaged over 200 repetitions).}
    \label{Table: the IS distribution averaged over 200 runs of BiCE for Example 5.2}
    \begin{tabular}{llllllll}
        \hline
        state & edge 14 & edge 16 & edge 17 & edge 4 & edge 13 & edge 19 & the rest\\
		\hline        
        0 & 0.334 & 0.170 & 0.179 & 0.128 & 0.123 & 0.127 & $\approx 0.002$ \\
        3 & 0.628 & 0.355 & 0.348 & 0.254 & 0.254 & 0.250 & $\approx 0.10$\\
        5 & 0.037 & 0.475 & 0.473 & 0.618 & 0.623 & 0.624 & $\approx 0.898$\\        
        \hline  
    \end{tabular}
\end{table} 
\subsection{Power transmission network with cascading failure}
\begin{figure}[!h]
    \centering
	\includegraphics[scale=0.5]{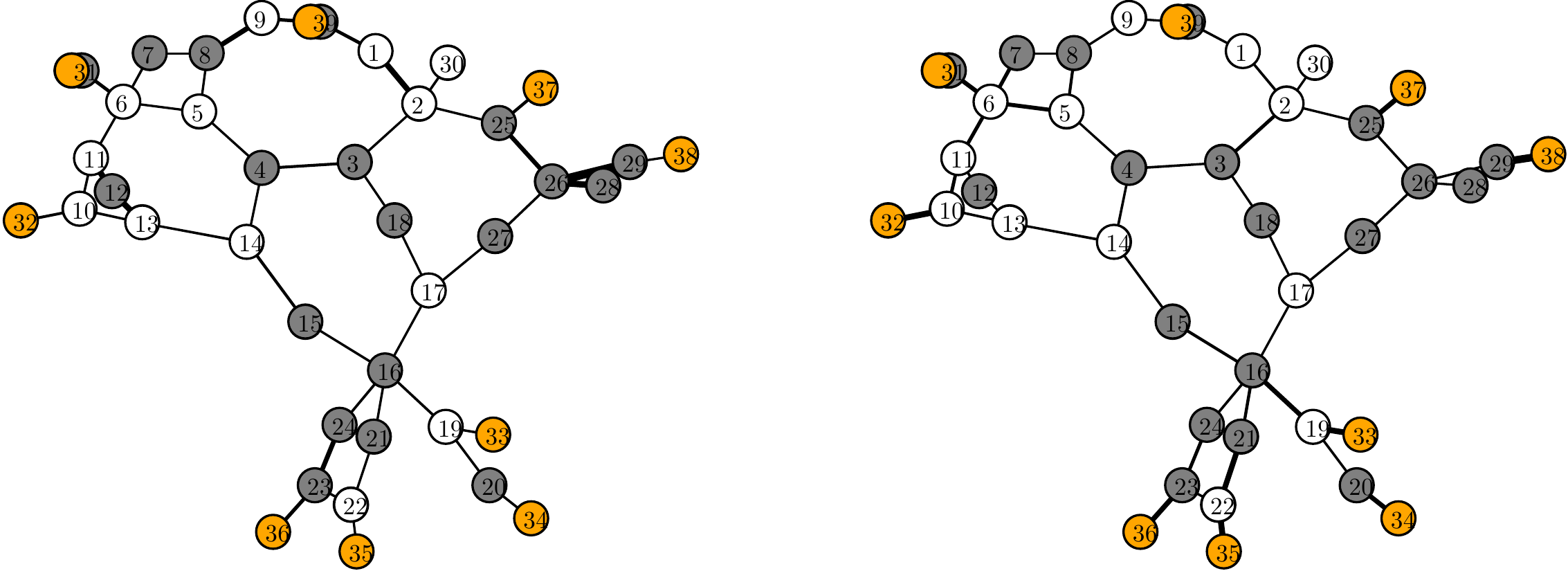}
	\caption{IEEE39 bus system, with edge thicknesses proportional to their capacities (left) and reactances (right).}
	\label{Fig: IEEE39 bus system}
\end{figure}

In this example, we consider the IEEE39 benchmark system, a simplified model of the high voltage transmission system in the northeast of the U.S.A. The model was first presented in 1970 \cite{Bills&others1970} and has been extensively used as a benchmark model in power system analysis \cite{Athay&others1979, Scherb&others2017, Rosero-Velasquez&Straub2019}.\\ 
It consists of 39 buses including 10 generators and 19 load buses, 34 transmission lines, and 12 transformers. The topology of the network is illustrated in Fig. \ref{Fig: IEEE39 bus system} where all the buses are modeled as nodes and transmission lines together with transformers are modeled as edges, so there are in total 39 nodes and 46 edges in the model. In the figure, orange circles stand for the source nodes, representing the 10 generators, and grey circles represent the terminal nodes, the 19 load buses. Edges are weighted by their reactance values shown on the right-hand side of Fig. \ref{Fig: IEEE39 bus system} and by their capacities shown on the left-hand side.\\ 
Through solving the direct current load flow (DCLF) problem described in the literature (e.g. \cite{Grainger1999} for IEEE39 benchmark model), one can derive the actual direct current (DC) that passes through each edge of the network. An edge fails when the DC flow exceeds its capacity, and the initial edge failures change the topology of the network resulting in a new configuration of the flow across the remaining components, which in turn may lead to further overloading of the edge. This phenomenon is also known as cascading failure and is modeled here based on \cite{Crucitti&others2004}. The system will finally reach an equilibrium state where no further edges are overloaded. In general, only a part of the original power demand at the load buses (the terminal nodes) can be matched in the equilibrium.\\
We assume that nodes will never fail, and the state of each edge follows an i.i.d Bernoulli distribution, with component failure probability $10^{-3}$. The LSF is then defined as a function of the system state $\bm{x}$, which is a binary vector, as follows:    
\begin{equation}
\label{Eq: LSF for power network system}
g_3(\bm{x})=30\% - L(\bm{x}),
\end{equation}
where $L(\bm{x})$ denotes the percentage of the original power demand that cannot be matched when the system achieves the equilibrium after the cascading failure, and the failure probability is defined as the probability of such percentage loss being greater or equal than the threshold, 30\%.\\
A crude MCS procedure with $10^6$ samples is used to validate the results of the proposed method, which gives a failure probability of $ 9.3 \cdot 10^{-5}$. We then apply the BiCE algorithm and set the hyperparameters $N=1,000, \delta_{tar}=\delta_{\epsilon}=1.5$. Two different kinds of prior distributions are considered here, the uninformative prior $\text{Beta}(\cdot; \bm{\theta}=[1,1]^T)$ and an informative prior $\text{Beta}(\cdot; \bm{\theta}=[10,10]^T)$, which correspond to setting $b=1$ and $b=10$, respectively. The second and third column of Table \ref{Table: performance of BiCE for Example 5.3} shows the performance of the BiCE methods after 500 repeated runs. The BiCE estimator with the uninformative prior performs better than the crude MCS with the same computational cost and also the standard iCE with $N=2,000$. Although standard iCE gives a smaller c.o.v. than BiCE with a uniform prior, the obtained estimate is strongly biased, which is likely due to the zero count problem. 
Moreover, the performance of the BiCE can be significantly enhanced through setting a larger $b$,e.g., 10. The sample c.o.v. of the BiCE with this informative prior $\text{Beta}(\cdot; \bm{\theta}=[10,10]^T)$ is 0.143, which is the smallest among all 4 cases. This setting also requires the lowest computational cost.
\begin{table}[!htbp]
    \centering
    \caption{Performance of BiCE for Example 5.3.}
    \label{Table: performance of BiCE for Example 5.3}
    \begin{tabular}{lllll}
        \hline
        method & BiCE & BiCE & MCS & iCE\\
        sample size, $N$ & $1,000$ & $1,000$ &  $5,746$ & $2,000$\\
        prior parm., $b$ & $1$ & $10$ & $\slash$ & $\slash$ \\
        \hline
        relative bias & 0.0049 & -0.009 & 0 & -0.409 \\
        sample c.o.v. & 0.5836& 0.143 & 1.367 & 0.305\\
        comp. cost & $5,746$ & $5,000$ & $5,746$ &  $11,870$\\
        \hline  
    \end{tabular}
\end{table} 
\section{Conclusions}
This paper studies the improved cross entropy (iCE) method in the context of network reliability assessment. We distinguish three distributions involved in iCE procedure: the optimal importance sampling (IS) distribution $p^*_{\bm{X}}(\bm{x})$, the suboptimal IS distribution $h(\bm{x}; \bm{v}^*)$, and the chosen IS distribution $h(\bm{x}; \widehat{ \bm{v} })$. Given a certain parametric family, the 'distance' between $p^*_{\bm{X}}(\bm{x})$ and $h(\bm{x}; \bm{v}^*)$ is fixed, and the objective of the CE method is to find a good estimator $\widehat{ \bm{v} }$ that is close to the optimal but inaccessible CE parameter $\bm{v}^*$. For parametric models that belong to the exponential family, $\widehat{ \bm{v} }$ is the self-normalized IS estimator of the optimal CE parameter $\bm{v}^*$, and hence converges to $\bm{v}^*$ as the sample size goes to infinity. Moreover, we show that $\hat{\bm{v}}$ can be viewed as the solution of a weighted maximum likelihood estimation (MLE) problem given the samples obtained at a certain level of the adaptive iCE sampling process. In network reliability assessments with discrete multi-state inputs, the parametric family can be chosen as the independent categorical distribution. In these approaches, the CE estimator $\widehat{ \bm{v} }$ suffers from the 'zero count problem', which is essentially the overfitting issue, resulting in a poor IS estimator with a strong negative bias. This paper derives the posterior predictive distribution $h(\bm{x}; \widetilde{ \bm{v} })$ to update the categorical model instead of the original maximum likelihood estimation $\hat{\bm{v}}$. By introducing the symmetric Dirichlet prior as shown in Eq.(\ref{Eq: a symmetric Dirichlet prior}), the probability assigned to each $d$-th category of the parametric model is at least $\frac{b}{b \cdot n_d+N}$, where $b$ is the hyper parameter. Hence, the 'zero count problem' is less likely to occur. The Bayesian estimator $\widetilde{ \bm{v} }$ is consistent, i.e., $\widetilde{ \bm{v} }$ converges to $\bm{v}^*$ as the sample size goes infinity. Combining the Bayesian estimator $\widetilde{ \bm{v} }$ with the standard iCE procedure, a modified iCE method called Bayesian improved cross entropy (BiCE) method is proposed for network reliability analysis. The efficiency and accuracy of the proposed method are illustrated through a set of numerical examples, from which it is found that BiCE with an appropriately chosen informative prior can significantly enhance the performance of the iCE method. Our numerical investigations indicate that a uniform prior performs only suboptimal. In all examples, we observed significantly better performance with a symmetric, informative prior with $b>1$, e.g., 5 or 10. It should also be stressed that the BiCE estimator can be skewed, and this is probably due to the limited capacity of the parametric model because of its assumption of independence. If the suboptimal IS distribution $h(\bm{x}; \bm{v}^*)$ itself is far from the optimal IS distribution $p^*_{\bm{X}}(\bm{x})$, independent on close $\widetilde{ \bm{v} }$ is to $\bm{v}^*$, the resulting IS estimator is bound to perform poorly 
\section{Acknowledgment}
The first author gratefully acknowledges the financial support of the China Scholarship Council.

\appendix
\section{Self-normalized importance sampling and cross entropy method}
\label{Appendix: Self-normalized importance sampling and cross entropy method}
In this Appendix, we first introduce the self-normalized IS estimator for estimating the expectation of a general function $H(\bm{x})$ and then prove that, in the CE (or iCE) method, the chosen parameter vector $\hat{\bm{v}}$ is the self normalized IS estimator of the sub-optimal parameter parameter vector $\bm{v}^*$ for the exponential parametric family.
\subsection{Self-normalized importance sampling}
We consider the following expectation of a general function $H(x)$:
\begin{equation}
\label{Eq: expectation with respect to unnormalized input distribution}
\mu = \mathbb{E}_{\pi}[H(\bm{X})].
\end{equation}
The input distribution $\pi(\bm{x})$ is only known pointwise to an unknown constant $Z$. That is
\begin{equation}
\label{Eq: unnormalized input distribution}
\pi(\bm{x}) = \frac{1}{Z}\pi_u(\bm{x}).
\end{equation}
$\pi_u(\bm{x})$ is the unnormalized form of $\pi$. In such case, the standard IS estimator of Eq.(\ref{Eq: IS estimator}) cannot be applied, since the likelihood ratio $L$, which is defined as the ratio of the input distribution $\pi(\bm{x})$ to the IS distribution $p_{IS}(\bm{x})$, is intractable. Instead, the following self-normalized IS estimator can be applied
\begin{equation}
\label{Eq: self normalized IS estimator}
\bar{\mu} = \sum_{k=1}^N \frac{W(\bm{x}_k)}{\sum_k W(\bm{x}_k)} H(\bm{x}_k),
\end{equation}
where $W(\bm{x}_k) \triangleq \frac{\pi_u(\bm{x}_k)}{p_{IS}(\bm{x}_k)}$. It can be proved that the self-normalized IS estimator is consistent, i.e., the estimator converges to the exact value as $N$ goes infinity, under the condition that the sample space of the input distribution $\pi(\bm{x})$ is included in that of the IS distribution $p_{IS}(\bm{x})$ \cite{Owen2013}. The estimator of Eq.(\ref{Eq: self normalized IS estimator}) can be less efficient than the standard Monte Carlo estimator that samples directly from $\pi(\bm{x})$.
The efficiency of the self-normalized estimator with respect to the crude Monte Carlo estimator can be measured by the so-called effective sample size (ESS) \cite{Kong1992}. ESS represents the number of samples that a crude MCS would need in order to yield the same variance as that of the self-normalized IS estimator of Eq.(\ref{Eq: self normalized IS estimator}). The ESS can be approximated through the following expression \cite{Kong1992}
\begin{equation}
\label{Eq: ESS}
ESS \approx \frac{N}{1+\delta^2(W(\bm{X}))}, \quad  \bm{X} \sim p_{IS}(\bm{x}),
\end{equation}
where $\delta(W(\bm{X}))$ represents the coefficient of variation of the weights $W(\bm{X})$ in Eq.(\ref{Eq: self normalized IS estimator}), and $N$ is the sample size of the self-normalized IS estimator of Eq.(\ref{Eq: self normalized IS estimator}).

\subsection{Cross entropy method with exponential parametric family}
In this subsection, we aim at finding a distribution from the exponential family $h(\bm{x};\bm{v})$ that has the minimal KL divergence with respect to the distribution $\pi$ of Eq.(\ref{Eq: unnormalized input distribution}). 
Note that the optimal IS distribution $p^*_{\bm{X}}(\bm{x})$ in Eq.(\ref{Eq: optimal IS distribution}) and the intermediate target distribution $p^{(t)}$ in Eq.(\ref{Eq: intermediate target distribution of iCE method}) (or Eq.(\ref{Eq: intermediate target distribution of CE method})) can be regarded as special cases of $\pi$, with $\pi_u$ set equal to $p_{\bm{X}}(\bm{x})\mathbb{I}\{ g(\bm{x}) \leq 0 \}$ and $p_{\bm{X}}(\bm{x})\Phi(-g(\bm{x})/\sigma^{(t)})$ (or $p_{\bm{X}}(\bm{x})\mathbb{I} \{ g(\bm{x}) \leq \gamma^{(t)} \}$), respectively.\\
The corresponding CE optimization problem is given as
\begin{equation}
\label{Eq: general CE optimization prob.}
\bm{v}^* = \argmax\limits_{\bm{v}\in\mathcal{V}} 
\sum_{\bm{x}\in \Omega_X} \pi_u(\bm{x})\ln(h(\bm{x}; \bm{v}))
\end{equation}
where $\Omega_{\bm{X}}$ is the sample space of $\bm{X}$. The summation in Eq.(\ref{Eq: general CE optimization prob.}) is substituted with the integral for continuous $\bm{X}$. The sample counterpart of Eq.(\ref{Eq: general CE optimization prob.}) reads 
\begin{equation}
\label{Eq: sample counterpart of general CE optimization prob.}
\widehat{\bm{v}} = \argmax\limits_{\bm{v}\in\mathcal{V}} \frac{1}{N} \sum \limits_{i=1}^{N} \frac{\pi_u(\bm{x})}{p_{ref}(\bm{x}_k)}\ln(h(\bm{x}_i; \bm{v})), \quad\quad \bm{x}_i \sim p_{ref}(\cdot).
\end{equation}\\
The exponential family of distributions is defined as the collection of distributions that have the following general form:
\begin{equation}
f(\bm{x}; \bm{\eta}) = a(\bm{x})\text{exp}(\bm{\eta}^{\text{T}}\bm{t}(\bm{x}) -A(\bm{\eta})),
\end{equation}
where $\bm{\eta}=(\eta_1,...,\eta_m)^{\text{T}}$ is often referred to as the canonical parameter. The statistic $\bm{t}(\bm{x})=(t_1(\bm{x}), ...,t_n(\bm{x}))^{\text{T}}$ is referred to as the sufficient statistic. The function $A(\bm{\eta})$ is known as the cumulant function.\\ 
In the following, we reparameterize the exponential family with $\bm{v} = \nabla _{\bm{\eta}}A(\bm{\eta})$. Through inserting $h(\bm{x}; \bm{v})$ into Eq.(\ref{Eq: general CE optimization prob.}) and setting the gradient of the objective function equal to zero, we get $\bm{v}^*_{c} = \mathbb{E}_{\pi}[\bm{t}(\bm{X})]$ \cite{Kroese&others2013}. Typically, $\bm{v}^*_{c}$ satisfies the constraint $\bm{v}\in\mathcal{V}$, and we have $\bm{v}^*=\bm{v}^*_{c}$. Note that the explicit expression of $\bm{v}^*$ depends on a prior knowledge of the distribution of $\pi(\bm{x})$, or equivalently a knowledge of the unknown constant $Z$ and therefore cannot be directly used. In such case, the sample counterpart Eq.(\ref{Eq: sample counterpart of general CE optimization prob.}) is solved instead, which gives us $\widehat{\bm{v}}=\frac{ \sum_{k=1}^{N} W(\bm{x_k})\bm{t}(\bm{x}_k) }{\sum_{k=1}^{N} W(\bm{x_k})}$, where $ W(\bm{x}_k) = \frac{\pi_u(\bm{x})}{p_{ref}(\bm{x}_k)}$.\\ 
According to Eq.(\ref{Eq: self normalized IS estimator}), $\widehat{\bm{v}}$ is the self-normalized estimator of $\bm{v}^*$ with $H(\bm{x})$ being the sufficient statistic $\bm{t}(\bm{x})$. The accuracy of the estimator $\widehat{\bm{v}}$ can be measured by the ESS defined in Eq.(\ref{Eq: ESS}).
 
\bibliographystyle{IeeeTran}
\bibliography{mybib}

\begin{thebibliography}{10}
\providecommand{\url}[1]{#1}
\csname url@samestyle\endcsname
\providecommand{\newblock}{\relax}
\providecommand{\bibinfo}[2]{#2}
\providecommand{\BIBentrySTDinterwordspacing}{\spaceskip=0pt\relax}
\providecommand{\BIBentryALTinterwordstretchfactor}{4}
\providecommand{\BIBentryALTinterwordspacing}{\spaceskip=\fontdimen2\font plus
\BIBentryALTinterwordstretchfactor\fontdimen3\font minus
  \fontdimen4\font\relax}
\providecommand{\BIBforeignlanguage}[2]{{%
\expandafter\ifx\csname l@#1\endcsname\relax
\typeout{** WARNING: IEEEtran.bst: No hyphenation pattern has been}%
\typeout{** loaded for the language `#1'. Using the pattern for}%
\typeout{** the default language instead.}%
\else
\language=\csname l@#1\endcsname
\fi
#2}}
\providecommand{\BIBdecl}{\relax}
\BIBdecl

\bibitem{Li&Liu2021}
J.~Li and W.~Liu, \emph{Lifeline engineering systems: Network reliability
  analysis and aseismic design}.\hskip 1em plus 0.5em minus 0.4em\relax
  Springer Nature, 2021.

\bibitem{Ball&others1995}
M.~O. Ball, C.~J. Colbourn, and J.~S. Provan, ``Network reliability,''
  \emph{Handbooks in operations research and management science}, vol.~7, pp.
  673--762, 1995.

\bibitem{Kumamoto&others1980}
H.~Kumamoto, K.~Tanaka, K.~Inoue, and E.~J. Henley, ``Dagger-sampling {M}onte
  {C}arlo for system unavailability evaluation,'' \emph{IEEE Transactions on
  Reliability}, vol.~29, no.~2, pp. 122--125, 1980.

\bibitem{Fishman1986a}
G.~S. Fishman, ``A {M}onte {C}arlo sampling plan for estimating network
  reliability,'' \emph{Operations Research}, vol.~34, no.~4, pp. 581--594,
  1986.

\bibitem{Fishman1989}
------, ``{M}onte {C}arlo estimation of the maximal flow distribution with
  discrete stochastic arc capacity levels,'' \emph{Naval Research Logistics
  (NRL)}, vol.~36, no.~6, pp. 829--849, 1989.

\bibitem{Alexopoulos&Fishman1991}
C.~Alexopoulos and G.~S. Fishman, ``Characterizing stochastic flow networks
  using the {M}onte {C}arlo method,'' \emph{Networks}, vol.~21, no.~7, pp.
  775--798, 1991.

\bibitem{Elperin&others1992}
T.~Elperin, I.~Gertsbakh, and M.~Lomonosov, ``An evolution model for {M}onte
  {C}arlo estimation of equilibrium network renewal parameters,''
  \emph{Probability in the Engineering and Informational Sciences}, vol.~6,
  no.~4, pp. 457--469, 1992.

\bibitem{Cancela&Khadiri2003}
H.~Cancela and M.~El~Khadiri, ``The recursive variance-reduction simulation
  algorithm for network reliability evaluation,'' \emph{IEEE Transactions on
  Reliability}, vol.~52, no.~2, pp. 207--212, 2003.

\bibitem{Rubino&Tuffin2009}
G.~Rubino and B.~Tuffin, \emph{Rare event simulation using {M}onte {C}arlo
  methods}.\hskip 1em plus 0.5em minus 0.4em\relax John Wiley \& Sons, 2009.

\bibitem{Zio2013}
E.~Zio, \emph{{M}onte {C}arlo simulation: The method}.\hskip 1em plus 0.5em
  minus 0.4em\relax Springer, 2013.

\bibitem{Behrensdorf&others2021}
J.~Behrensdorf, T.-E. Regenhardt, M.~Broggi, and M.~Beer, ``Numerically
  efficient computation of the survival signature for the reliability analysis
  of large networks,'' \emph{Reliability Engineering \& System Safety}, vol.
  216, p. 107935, 2021.

\bibitem{Zio&Pedroni2008}
E.~Zio and N.~Pedroni, ``Reliability analysis of discrete multi-state systems
  by means of subset simulation,'' in \emph{Proceedings of the 17th ESREL
  Conference}, 2008, pp. 22--25.

\bibitem{Botev&others2013}
Z.~I. Botev, P.~L'Ecuyer, G.~Rubino, R.~Simard, and B.~Tuffin, ``Static network
  reliability estimation via generalized splitting,'' \emph{INFORMS Journal on
  Computing}, vol.~25, no.~1, pp. 56--71, 2013.

\bibitem{Zuev&others2015}
K.~M. Zuev, S.~Wu, and J.~L. Beck, ``General network reliability problem and
  its efficient solution by subset simulation,'' \emph{Probabilistic
  Engineering Mechanics}, vol.~40, pp. 25--35, 2015.

\bibitem{Botev&others2018}
Z.~I. Botev, P.~l'Ecuyer, and B.~Tuffin, ``Reliability estimation for networks
  with minimal flow demand and random link capacities,'' \emph{arXiv preprint
  arXiv:1805.03326}, 2018.

\bibitem{Jensen&Jerez2018}
H.~A. Jensen and D.~J. Jerez, ``A stochastic framework for reliability and
  sensitivity analysis of large scale water distribution networks,''
  \emph{Reliability Engineering \& System Safety}, vol. 176, pp. 80--92, 2018.

\bibitem{Chan&others2022a}
J.~Chan, I.~Papaioannou, and D.~Straub, ``An adaptive subset simulation
  algorithm for system reliability analysis with discontinuous limit states,''
  \emph{Reliability Engineering \& System Safety}, p. 108607, 2022.

\bibitem{Bulteau&Khadiri2002}
S.~Bulteau and M.~El~Khadiri, ``A new importance sampling {M}onte {C}arlo
  method for a flow network reliability problem,'' \emph{Naval Research
  Logistics (NRL)}, vol.~49, no.~2, pp. 204--228, 2002.

\bibitem{Hui&others2003}
K.-P. Hui, N.~Bean, M.~Kraetzl, and D.~Kroese, ``Network reliability estimation
  using the tree cut and merge algorithm with importance sampling,'' in
  \emph{Proceedings of the 4th International Workshop on Design of Reliable
  Communication Networks}.\hskip 1em plus 0.5em minus 0.4em\relax IEEE, 2003,
  pp. 254--262.

\bibitem{Wang&others2016}
Z.~Wang and J.~Song, ``Cross-entropy-based adaptive importance sampling using
  von mises-fisher mixture for high dimensional reliability analysis,''
  \emph{Structural Safety}, vol.~59, pp. 42--52, 2016.

\bibitem{Dehghani&others2021}
N.~L. Dehghani, S.~Zamanian, and A.~Shafieezadeh, ``Adaptive network
  reliability analysis: Methodology and applications to power grid,''
  \emph{Reliability Engineering \& System Safety}, vol. 216, p. 107973, 2021.

\bibitem{Rubinstein1997}
R.~Y. Rubinstein, ``Optimization of computer simulation models with rare
  events,'' \emph{European Journal of Operational Research}, vol.~99, no.~1,
  pp. 89--112, 1997.

\bibitem{Papaioannou&others2019}
I.~Papaioannou, S.~Geyer, and D.~Straub, ``Improved cross entropy-based
  importance sampling with a flexible mixture model,'' \emph{Reliability
  Engineering \& System Safety}, vol. 191, p. 106564, 2019.

\bibitem{Murphy2012}
K.~P. Murphy, \emph{Machine learning: A probabilistic perspective}.\hskip 1em
  plus 0.5em minus 0.4em\relax MIT press, 2012.

\bibitem{Hui&others2005}
K.-P. Hui, N.~Bean, M.~Kraetzl, and D.~P. Kroese, ``The cross-entropy method
  for network reliability estimation,'' \emph{Annals of Operations Research},
  vol. 134, no.~1, p. 101, 2005.

\bibitem{Elperin&others1991}
T.~Elperin, I.~Gertsbakh, and M.~Lomonosov, ``Estimation of network reliability
  using graph evolution models,'' \emph{IEEE Transactions on Reliability},
  vol.~40, no.~5, pp. 572--581, 1991.

\bibitem{Li&He2002}
J.~Li and J.~He, ``A recursive decomposition algorithm for network seismic
  reliability evaluation,'' \emph{Earthquake engineering \& structural
  dynamics}, vol.~31, no.~8, pp. 1525--1539, 2002.

\bibitem{Hardy&others2007}
G.~Hardy, C.~Lucet, and N.~Limnios, ``K-terminal network reliability measures
  with binary decision diagrams,'' \emph{IEEE Transactions on Reliability},
  vol.~56, no.~3, pp. 506--515, 2007.

\bibitem{Paredes&others2019}
R.~Paredes, L.~Due{\~n}as-Osorio, K.~S. Meel, and M.~Y. Vardi, ``Principled
  network reliability approximation: A counting-based approach,''
  \emph{Reliability Engineering \& System Safety}, vol. 191, p. 106472, 2019.

\bibitem{Miao&others2020}
H.~Miao, W.~Liu, and J.~Li, ``Seismic reliability analysis of water
  distribution networks on the basis of the probability density evolution
  method,'' \emph{Structural Safety}, vol.~86, p. 101960, 2020.

\bibitem{Byun&Song2021}
J.-E. Byun and J.~Song, ``Generalized matrix-based bayesian network for
  multi-state systems,'' \emph{Reliability Engineering \& System Safety}, vol.
  211, p. 107468, 2021.

\bibitem{Owen2013}
A.~B. Owen, \emph{{M}onte {C}arlo theory, methods and examples}.\hskip 1em plus
  0.5em minus 0.4em\relax Standford, 2013.

\bibitem{Madsen&others2006}
H.~O. Madsen, S.~Krenk, and N.~C. Lind, \emph{Methods of structural
  safety}.\hskip 1em plus 0.5em minus 0.4em\relax Courier Corporation, 2006.

\bibitem{Au&Beck1999}
S.-K. Au and J.~L. Beck, ``A new adaptive importance sampling scheme for
  reliability calculations,'' \emph{Structural safety}, vol.~21, no.~2, pp.
  135--158, 1999.

\bibitem{Rubinstein&Kroese2016}
R.~Y. Rubinstein and D.~P. Kroese, \emph{Simulation and the {M}onte {C}arlo
  method}.\hskip 1em plus 0.5em minus 0.4em\relax John Wiley \& Sons, 2016,
  vol.~10.

\bibitem{Au&Beck2003}
S.-K. Au and J.~Beck, ``Important sampling in high dimensions,''
  \emph{Structural safety}, vol.~25, no.~2, pp. 139--163, 2003.

\bibitem{Boyd&Vandenberghe2004}
S.~Boyd and L.~Vandenberghe, \emph{Convex optimization}.\hskip 1em plus 0.5em
  minus 0.4em\relax Cambridge university press, 2004.

\bibitem{Chan&Kroese2012}
J.~C. Chan and D.~P. Kroese, ``Improved cross-entropy method for estimation,''
  \emph{Statistics and computing}, vol.~22, no.~5, pp. 1031--1040, 2012.

\bibitem{Geyer&others2019}
S.~Geyer, I.~Papaioannou, and D.~Straub, ``Cross entropy-based importance
  sampling using gaussian densities revisited,'' \emph{Structural Safety},
  vol.~76, pp. 15--27, 2019.

\bibitem{Kroese&others2013}
D.~P. Kroese, T.~Taimre, and Z.~I. Botev, \emph{Handbook of {M}onte {C}arlo
  methods}.\hskip 1em plus 0.5em minus 0.4em\relax John Wiley \& Sons, 2013,
  vol. 706.

\bibitem{Uribe&others2021}
F.~Uribe, I.~Papaioannou, Y.~M. Marzouk, and D.~Straub, ``Cross-entropy-based
  importance sampling with failure-informed dimension reduction for rare event
  simulation,'' \emph{SIAM/ASA Journal on Uncertainty Quantification}, vol.~9,
  no.~2, pp. 818--847, 2021.

\bibitem{Chan&others2022b}
J.~Chan, I.~Papaioannou, and D.~Straub, ``Improved cross entropy-based
  importance sampling for network reliability assessment,'' in
  \emph{Proceedings of the 13th International Conference on Structural Safety
  \& Reliability}.\hskip 1em plus 0.5em minus 0.4em\relax ICOSSAR, 2022.

\bibitem{Ramirez-marquez&Coit2005}
J.~E. Ramirez-Marquez and D.~W. Coit, ``A monte-carlo simulation approach for
  approximating multi-state two-terminal reliability,'' \emph{Reliability
  Engineering \& System Safety}, vol.~87, no.~2, pp. 253--264, 2005.

\bibitem{Bills&others1970}
G.~Bills, ``On-line stability analysis study, rp 90-1,'' North American
  Rockwell Information Systems Co., Anaheim, CA (USA), Tech. Rep., 1970.

\bibitem{Athay&others1979}
T.~Athay, R.~Podmore, and S.~Virmani, ``A practical method for the direct
  analysis of transient stability,'' \emph{IEEE Transactions on Power Apparatus
  and Systems}, no.~2, pp. 573--584, 1979.

\bibitem{Scherb&others2017}
A.~Scherb, L.~Garr{\`e}, and D.~Straub, ``Reliability and component importance
  in networks subject to spatially distributed hazards followed by cascading
  failures,'' \emph{ASCE-ASME Journal of Risk and Uncertainty in Engineering
  Systems, Part B: Mechanical Engineering}, vol.~3, no.~2, 2017.

\bibitem{Rosero-Velasquez&Straub2019}
H.~Rosero-Vel\'{a}squez and D.~Straub, ``Representative natural hazard
  scenarios for risk assessment of spatially distributed infrastructure
  systems,'' in \emph{The 29th European Safety and Reliability Conference},
  2019, pp. 1--7.

\bibitem{Grainger1999}
J.~J. Grainger, \emph{Power system analysis}.\hskip 1em plus 0.5em minus
  0.4em\relax McGraw-Hill, 1999.

\bibitem{Crucitti&others2004}
P.~Crucitti, V.~Latora, and M.~Marchiori, ``Model for cascading failures in
  complex networks,'' \emph{Physical Review E}, vol.~69, no.~4, p. 045104,
  2004.

\bibitem{Kong1992}
A.~Kong, ``A note on importance sampling using standardized weights,''
  \emph{University of Chicago, Dept. of Statistics, Tech. Rep}, vol. 348, 1992.

\end{thebibliography}






\end{document}